\pdfoutput=1
\documentclass[12pt,preprint]{emulateapj}
\usepackage{amssymb,mathrsfs,color}
\usepackage{amsmath}
\newfont{\rsfsten}{rsfs10 scaled 1200}
\newfont{\rsfsseven}{rsfs7 scaled 1200}
\newfont{\rsfsfive}{rsfs5 scaled 1200}

\slugcomment{First draft: May 31, 2013; submitted: September 9, 2015; accepted \today}

\def\Alfvenic{Alfv$\acute{\rm e}$nic~}
\shorttitle{The Radio Bright Zone at The Galactic Center}
\shortauthors{Zhao, Morris \& Goss}

\begin{document}
\title{A New Perspective of the Radio Bright Zone at The Galactic Center: Feedback from Nuclear Activities}
\author{
Jun-Hui Zhao\altaffilmark{1}, Mark R. Morris\altaffilmark{2} \& W. M. Goss\altaffilmark{3}
}
\affil{$^1$Harvard-Smithsonian Center for Astrophysics, 60
Garden Street, Cambridge, MA 02138, USA; jzhao@cfa.harvard.edu}
\affil{$^2$Department of Physics and Astronomy, University of California Los Angeles, Los Angeles, CA 90095}
\affil{$^3$NRAO, P.O. Box O, Socorro, NM 87801, USA}

\begin{abstract}
New observations of Sgr A have been carried out with the Jansky VLA in 
the B and C arrays using the broadband (2 GHz) continuum mode at 5.5 GHz. 
The field of view covers the central 13\arcmin~(30 pc) region of the 
radio-bright zone at the Galactic center. Using the multi-scale and 
multi-frequency-synthesis (MS-MFS) algorithms in CASA, we have imaged 
Sgr A with a resolution of 1\arcsec, achieving an rms noise of 8 $\mu$Jy 
beam$^{-1}$, and a dynamic range of 100,000:1. Both previously known 
and newly identified radio features in this region are revealed, 
including numerous filamentary sources. The radio continuum image 
is compared with Chandra X-ray images, with a CN emission-line 
image obtained with the SMA and with detailed Paschen-$\alpha$ images 
obtained with HST/NICMOS. We discuss several prominent features in 
the radio image. The ``Sgr A West Wings'' extend 2\arcmin\,(5 pc) 
from the NW and SE tips of the Sgr A West HII region (the "Mini-spiral") 
to positions located 2.9 and 2.4 arc min to the northwest and 
southeast of Sgr A*, respectively. The NW wing, along  with several 
other prominent features, including the previously identified ``NW Streamers'', 
form an elongated radio lobe (NW lobe), oriented nearly perpendicular 
to the Galactic plane. This radio lobe, with a size of 
6.3\arcmin$\times$3.2\arcmin\,(14.4 pc$\times$7.3 pc), has a known 
X-ray counterpart. In the outer region of the NW lobe, a row of three 
thermally emitting rings is observed. A field containing numerous 
amorphous radio blobs extends for a distance of $\sim$ 2 arc min 
beyond the tip of the SE wing; these newly recognized features 
coincide with the SE X-ray lobe.  Most of the amorphous radio blobs 
in the NW and SE lobes have Paschen-$\alpha$ counterparts.  
We propose that they have been produced by shock interaction of 
ambient gas concentrations with a collimated nuclear wind or an 
outflow that originated from within the circumnuclear disk (CND). 
We also discuss the possibility that the ionized wind or outflow has 
been launched by radiation force produced by the central star cluster.
Finally, we remark on the detailed structure of a prominent radio emission 
feature located within the shell of the Sgr A East supernova remnant. Because
this feature -- the ``Sigma Front'' -- correlates well in shape and orientation with
the nearby edge of the CND, we propose that it is a reflected shock wave 
resulting from the impact of the Sgr A East blast wave on the CND.  
\end{abstract}

\section{Introduction}

The prominent $\gamma-$ray structures of $\sim$10 kpc scale, referred 
to as the Fermi bubbles \citep {su10},  are likely evidence of past AGN or 
starburst  activity at the Galactic Center.  Recent radio observations 
of the two giant, linearly polarized radio lobes with the Parkes telescope 
\citep{carr13} also show evidence of large scale manifestations of activity
at the Galactic Center. The origin of the Fermi bubbles with a luminosity 
of $4\times 10^{37}$ erg s$^{-1}$ and a total energy budget of 
$\sim$10$^{55}$ erg has been discussed and reviewed by \cite{pont13}. 
Although Sgr A* is presently in an extremely sub-Eddington accretion 
state, it can apparently undergo dramatic increases in its luminosity 
\citep{pont13},  thereby powering large-scale structures centered 
on it.  

These features may well originate from a starburst phase in the Galactic 
center within the past few million years \citep{su10} or to an intense 
AGN phase on the same time scale, with Sgr A* accreting near its Eddington 
limit \citep{zubo11}. Whatever the mechanism for producing the Fermi 
bubbles, the inferred energy injection rate is consistent with the 
IR luminosity and $\gamma$-ray flux emerging from the central molecular 
zone (CMZ) \citep{croc11}. 

Within the inner 40 parsecs (17\arcmin), a remarkable bipolar X-ray 
lobe structure ($\sim$10 pc) has been revealed in deep Chandra X-ray 
images \citep{bag03,morr03}; these lobes are aligned with the Fermi 
bubbles, but on an angular scale that is $\sim$200 times smaller than
that of the bubbles. \cite{morr03} point out that a number of X-ray 
clumps detected in the bipolar lobes  suggest a series of recurrent 
ejections from within the circumnuclear disk (CND), occurring on 
time scales of hundreds to thousands of years, depending on the 
velocities of the emitting clumps. 

Although it has not been clear how the Fermi-bubbles can be related
to past AGN and starburst activities in the Galactic center, 
the current activities from massive star formation produce significant 
feedback to the ISM in the central 40 parsecs via stellar winds as 
well as an intense UV radiation field originating from massive stars 
and supernova (SN). The SN of Sgr A East alone produces an output 
energy of 10$^{52-53}$ erg \citep{mezg89,mezg96}, substantially 
impacting the interstellar medium near Sgr A*, although the nature, 
location and environment of Sgr A East have not been fully determined. 
At 90 cm, the appearance of the Sgr A West Mini-spiral features in 
absorption against Sgr A East, conclusively shows that the former 
structure is situated in front of or is mixed with Sgr A East 
\citep{pedl89,goss89}. Based on the kinematics of maser spots 
associated with the SNR \citep{yusf99}, the Sgr A East SN is 
located at a distance $\le$ 5 pc from Sgr A*. If the central cluster 
with an IR luminosity of 4$\times10^7$ L$_\odot$ provides the heating 
of the SN dust in Sgr A East \citep{lau15}, best-fit models show 
the distance between the dust and Sgr A* to  vary between 3.6 to 
7.3 pc depending on the relevant physical parameters used in
the model of \cite{lau15}. The development and expansion of 
the Sgr A East shell (SNR) appear to be in a non-uniform density 
medium given the fact that a complex distribution of the +50
km s$^{-1}$ GMC, dense molecular clumps in the CND and velocity 
shears have been observed in the Galactic center. Interactions between 
the molecular clouds with the Sgr A East SNR \citep{eker83,sera92} 
along with the presence of velocity shears \citep[{\it e.g.}{\rm}]{uchi98} 
have been thought to be the cause of the apparent shape of the SNR 
shell that appears to be elongated in the Galactic longitude.

In this paper, in section 2 we summarize the setup of the Jansky 
Very Large Array (JVLA) observations. In section 3, we describe 
the new results conerning the central 13\arcmin\/, the radio bright 
zone (hereafter, {\bf RBZ}) at the Galactic center region at 5.5 GHz. 
This description follows the radio detection of the Cannonball 
from the new 2 GHz-bandwidth images that has been reported 
earlier \citep{zhao13}. In section 4, the astrophysical implications
of the radio observations are discussed.

\begin{table}[b]
\tablenum{1}
\setlength{\tabcolsep}{1.7mm}
\caption{Log of Observations}
\begin{tabular}{lcccl}
\hline\hline \\
{Observing date~~}&
{Array}&
{~~~$\nu$~~~} &
{~~~$\Delta\nu$~~~} &
\multicolumn{1}{c}{Calibrators}\\
{} &
{} &
{GHz} &
{GHz} &
\\
\hline
2012-03-29& C &5.5&2&J1331+3030 ({\it DB})\\
          &   &   & &J1733$-$1304 ({\it CG})\\
          &   &   & &J1331+3030 ({\it FD})\\
2012-04-22& C &5.5&2&J0319+4130 ({\it DB})\\
          &   &   & &J1733$-$1304 ({\it CG})\\
          &   &   & &J0137+3309 ({\it FD})\\
2012-07-24& B &5.5&2&J1331+3030 ({\it DB})\\
          &   &   & &J1733$-$1304 ({\it CG})\\
          &   &   & &J1331+3030 ({\it FD})\\
2012-07-27& B &5.5&2&J0319+4130 ({\it DB})\\
          &   &   & &J1733$-$1304 ({\it CG})\\
          &   &   & &J0137+3309 ({\it FD})\\
\hline
\end{tabular}\\
\begin{tabular}{p{0.45\textwidth}}
{\footnotesize 
For all the four epoch observations,
the target field was centered at
$\alpha_{\rm J2000}=17^{h}45^{m}42^{s}.718$~
$\delta_{\rm J2000}= -29$\arcdeg00\arcmin17\arcsec.97, 
and the default correlator setup (subbands$\times$channels=16$\times$1024)
for broadband continnum observations is used. The calibrator codes stand 
for delay \& bandpass ({\it DB}), complex gain ({\it CG}) and flux density ({\it FD}) scale,
respectively.}
\\
\end{tabular}
\end{table}
\section{Observations}

The JVLA observations of the Galactic center at 6 cm were carried out on 
March 29, April 22, July 24 and July 27 of 2012 in the C and B configurations, 
positioned near the geometrical center of the Sgr A East shell, offset by 
($\Delta\alpha=$35\arcsec, $\Delta\delta=$10\arcsec) from Sgr A*. Observations 
in each array configuration were scheduled with two four-hour observing blocks 
to cover a complete eight-hour uv-track with two separate observing dates. 
A standard continuum observing mode having a total bandwidth of 2 GHz in
16 subbands of 64 channels each was used for all four epochs of observations.  
The spectral resolution of each subband was 2 MHz. Table 1 summarizes the 
log of the JVLA observing program AZ198 or 12A-037. The details of the
data reduction for broadband imaging are discussed in Appendix A.
\begin{figure}[t]
\centering
\includegraphics[angle=0,width=102mm]{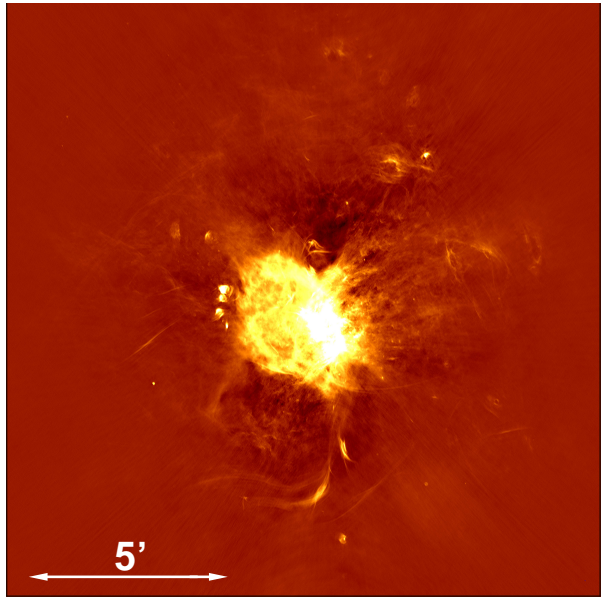}
\caption{
Overall radio structure - the 2012 JVLA C+B-array image of radio intensity 
at 5.5 GHz, constructed with uniform weighting and cleaned with MFS-MS 
algorithm,  showing the central 15\arcmin\/radio bright zone (RBZ) at the 
Galactic center. The rms noise is 8 $\mu$Jy beam$^{-1}$ with a FWHM beam 
(1.6\arcsec$\times$0.6\arcsec, PA=11\arcdeg).
}
\end{figure}

Owing to the dramatic recent improvements in the Very Large Array,
including hardware and software, we have obtained much deeper images 
than have heretofore been possible, achieving a dynamic range of 
100,000:1.

\section{Radio Bright Zone at the Galactic center}

The new 2-GHz broadband images at 5.5 GHz reveal additional  
fine structure and faint sources in the RBZ, compared to our previous 
VLA images \citep{zhao09} as well as those of many other investigators, 
{\it e.g.} \citep{eker83,lo83,ho85,yusf87a,pedl89}. In \cite{zhao13}, 
we have presented preliminary results of these new observations in 
a publication describing the radio counterpart of the X-ray ``Cannonball'', 
a pulsar wind nebula surrounding the neutron star that was apparently 
launched in the explosion that created the Sgr A East supernova remnant.  
In addition, these data were used to report our imaging of the radio 
filaments (Northern Filaments) of the NuSTAR source G359.97-0.038  \citep{nynk15}.

\subsection{Overall Radio Structure}

Fig. 1 shows the overall structure of the radio emission
from the Sgr A Complex  \citep{pedl89} observed at 5.5 GHz.  
In addition to the well-known sources such as Sgr A*, Sgr A West
and Sgr A East \citep{eker83}, the continuum image contains 
many prominent features, showing the details of the complex 
filamentary structure. As many as 129 compact radio sources 
outside of Sgr A East have been detected at 6 cm with peak 
flux densities ranging between 5 and 0.06 mJy beam$^{-1}$ 
(8$\sigma$) (ZMG 2015 in preparation).

\begin{figure*}[t]
\centering
\includegraphics[angle=270,width=200mm]{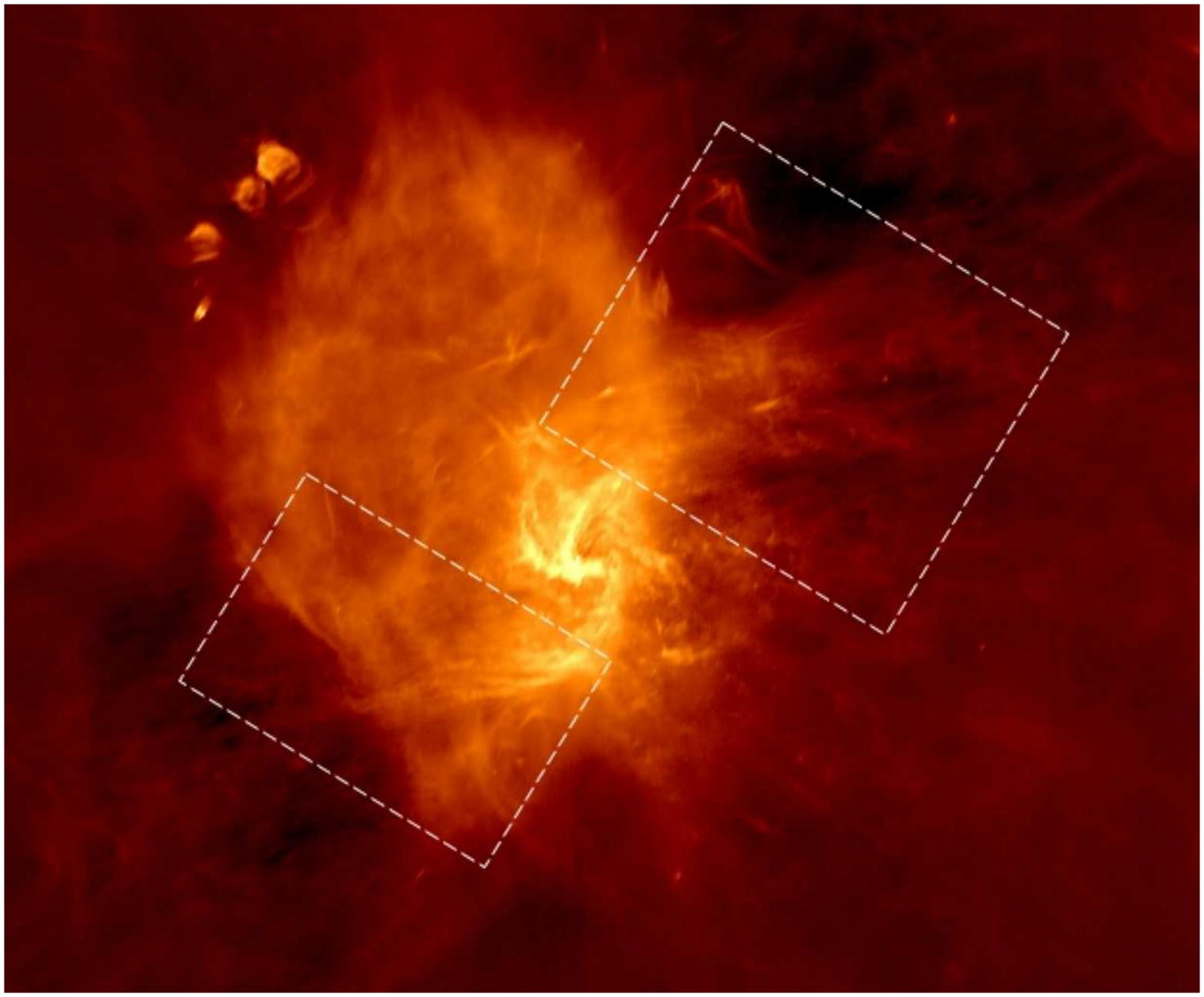} \\
\vskip -50pt
\includegraphics[angle=0,width=99.5mm]{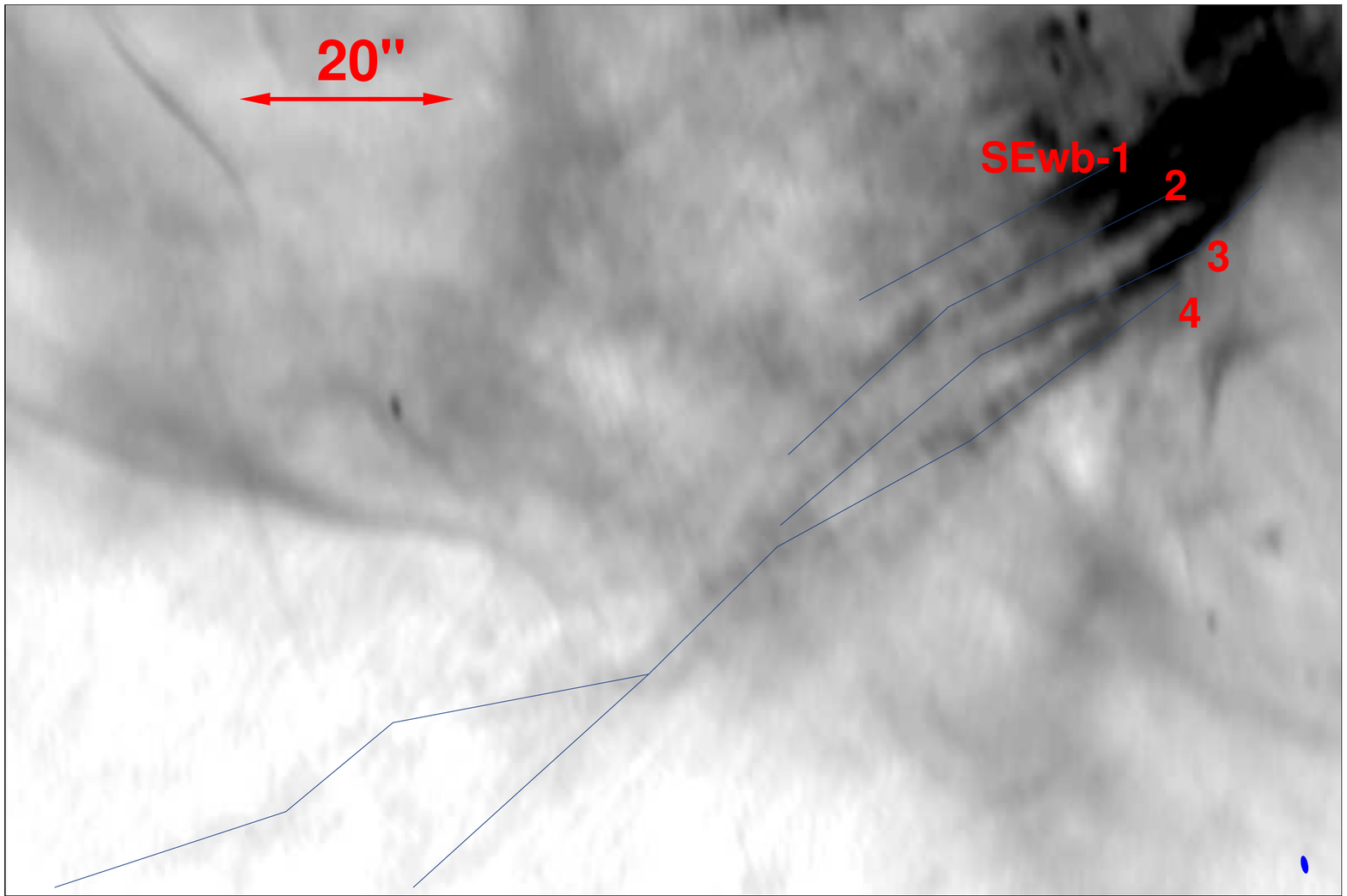} 
\includegraphics[angle=0,width=75mm]{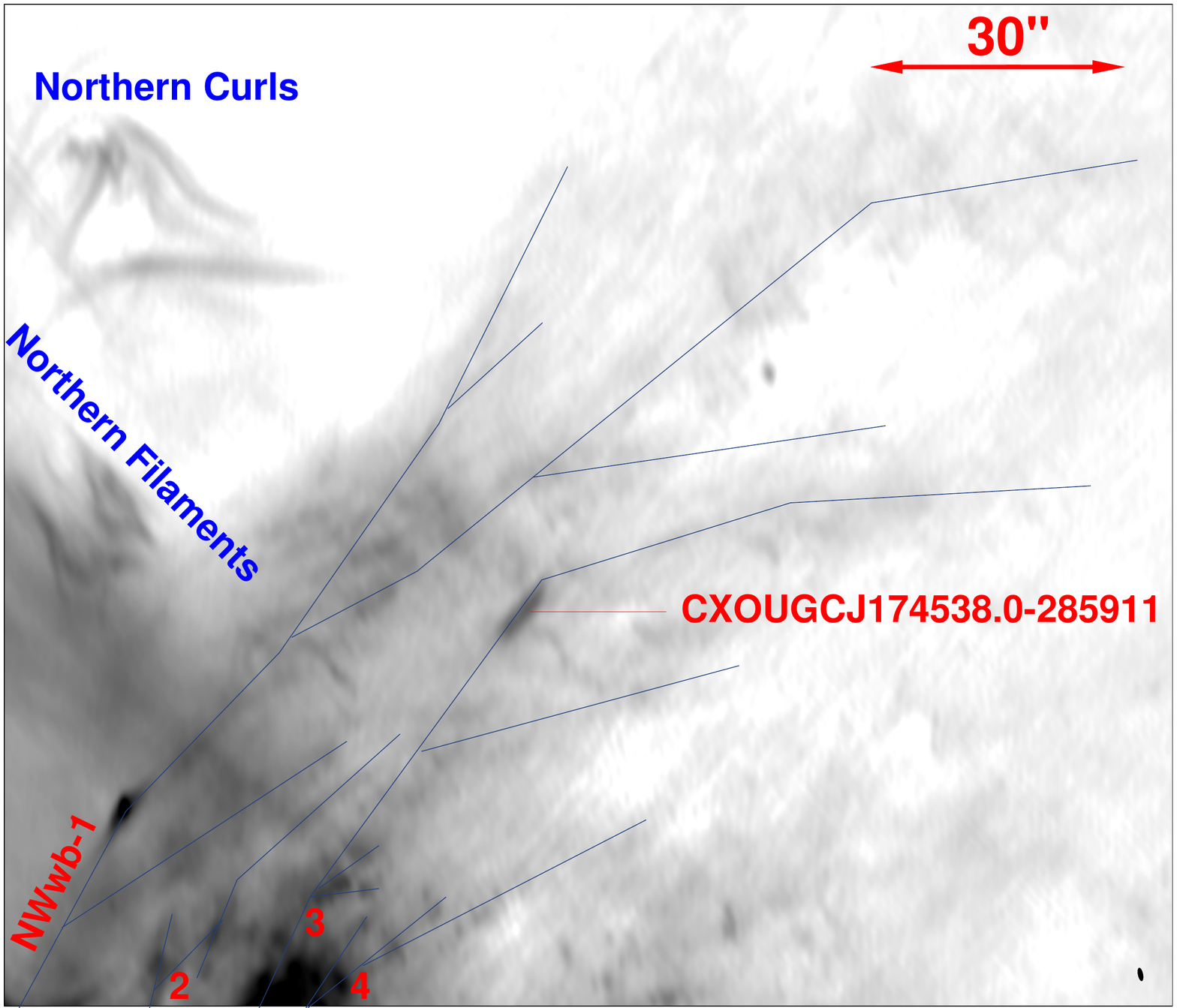}
\vskip 20pt
\caption{Top: a 5.5 GHz image in Galactic coordinates (increasing Galactic 
longitude is up) showing with dashed rectangles the regions containing the 
SE (lower-left) and NW (upper-right) "wings" that extend from the Mini-spiral 
arms, in the direction  nearly perpendicular to the Galactic plane. Bottom: 
subimages of the SE (left) and NW (right) wings in Equatorial coordinates 
(North up) corresponding to the SE and NW dashed rectangles in the top image. 
The FWHM beam (1.6\arcsec$\times$0.6\arcsec, PA=11\arcdeg) is shown at the 
bottom right of each subimage, The faint blue lines track the continuous 
branchings of the extended wings. The bright short filament (NW Streak), 
a radio counterpart of the compact X-ray source, CXOUGCJ174538.0-285911, 
is labelled. }
\end{figure*}

Within the shell of Sgr A East and the region near Sgr A*, 
numerous filamentary features are revealed in great detail.  
Their properties and astrophysical implications will be discussed 
in subsequent papers.

\subsection{Extended Wings of the Mini-spiral}

Bright thermal emission from Sgr A West -- the ``{\bf Mini-spiral}'' 
HII structure -- dominates the 5.5 GHz image.  As has been well established, 
{\it e.g.} \citep{morr96}, Sgr A West has three bright components lying 
inside the CND -- the Northern Arm, Eastern Arm and the Bar -- and 
its free-free emission also outlines the inner ionized edge of the CND, 
particularly toward the West, where it manifests as the Western Arc 
(see, e.g., Fig. 3). \cite{yusf87a}, also observing with the VLA at 6 cm, 
noted that emission from the arms of Sgr A West extends out of the interior 
of the CND forming radio ``streamers'' oriented perpendicular to the Galactic 
plane (see their Figs. 4 \& 5).

These extended emission features are particularly well characterized in 
our image (Figs. 2, 3, 5, and 6). They appear to be, in projection, 
continuations of the Mini-spiral arms, stretching linearly up to 
$\sim$3\arcmin~(7 pc in projection) along a position angle oriented 
at about +10 degrees in projection relative to the Galaxy's rotation 
axis, or about +20 degrees relative to the normal to the CND 
\citep{gust87,jack93,zhao10}. We hereafter refer to these emission 
structures as the ``{\bf NW and SE Wings}''. The SE Wing  
(Fig. 2, bottom-left) appears to split into several branches. We identify 
four of them (SEb1-4), each of which is indicated by line segments in Fig. 2. 
The peak intensity of the SE wing branches is several mJy beam$^{-1}$. 
They are slightly resolved, with a typical width of 2\arcsec\/. The length 
of SE wing branches ranges from 0.5\arcmin~to 2\arcmin. The longest one 
(SEb-4) appears to  bifurcate further into two sub-branches in the middle,
crossing the boundary of the Sgr A East shell. The bundle of SE wing filaments 
is clearly present in absorption at 90 cm against the Sgr A East shell 
\citep{pedl89,goss89}, indicating that the SE Wing becomes optically 
thick at the lower frequency and is situated in front of, or is spatially 
mixed with, the Sgr A East shell. In addition, in the region around 
the SE Wing, a few compact radio components were detected. The radio 
properties of these sources are summarized in Table 2.

The bottom image in Fig. 2 (left for SE and right for NW) shows the 
region containing the NW Wing, which appears to split into more branches 
than the SE Wing. Four of them, with emission intensity similar to 
that of the SE Wing, are labelled. The lengths of the NW wing branches 
range from $0.8$\arcmin~(NWb-2) to 3\arcmin~(NWb-1), and they appear 
to bifurcate to more sub-branches with increasing distance from Sgr A 
West. The radio properties of NW wing branches are summarized in Table 2.

\cite{yusf87a} and \cite{pedl89} pointed out the presence of an extended 
ionized halo around Sgr A East and West having a 90 cm optical depth of 
$\tau_{\rm 90}\sim3$ (or an emission measure of $\sim2\times10^5$ 
pc cm$^{-6}$). Both the SE and NW wing features are present as 
$\tau_{\rm 90}>3$ features in the map of the optical-depth distribution 
across Sgr A East derived from VLA 6 and 90 cm data, c.f., Fig. 10 of  
\cite{pedl89}, indicating that the emission at 90 cm is attenuated by 
the ionized gas in Sgr A West. From the 90 cm optical depth, and using 
the angular width of a typical wing branch ($\theta_{\rm w}\sim2\arcsec$), 
the electron density in the wing can be inferred to be
$$n_{\rm e} \lesssim 2\times10^3~{\rm cm}^{-3} 
\left(\theta_{\rm w}\over2\arcsec\right)^{-0.5}
\left(T_{\rm e} \over 10^4 {\rm K}\right)^{0.675}.$$
The electron density of the ionized gas in the wings therefore appears 
to be 1-2 orders of magnitude smaller than the high-density ionized 
clumps observed in Sgr A West \citep{zhao10}.

\begin{figure*}[t]
\centering
\hskip 0pt
\vskip -50pt
\includegraphics[angle=0,width=190mm]{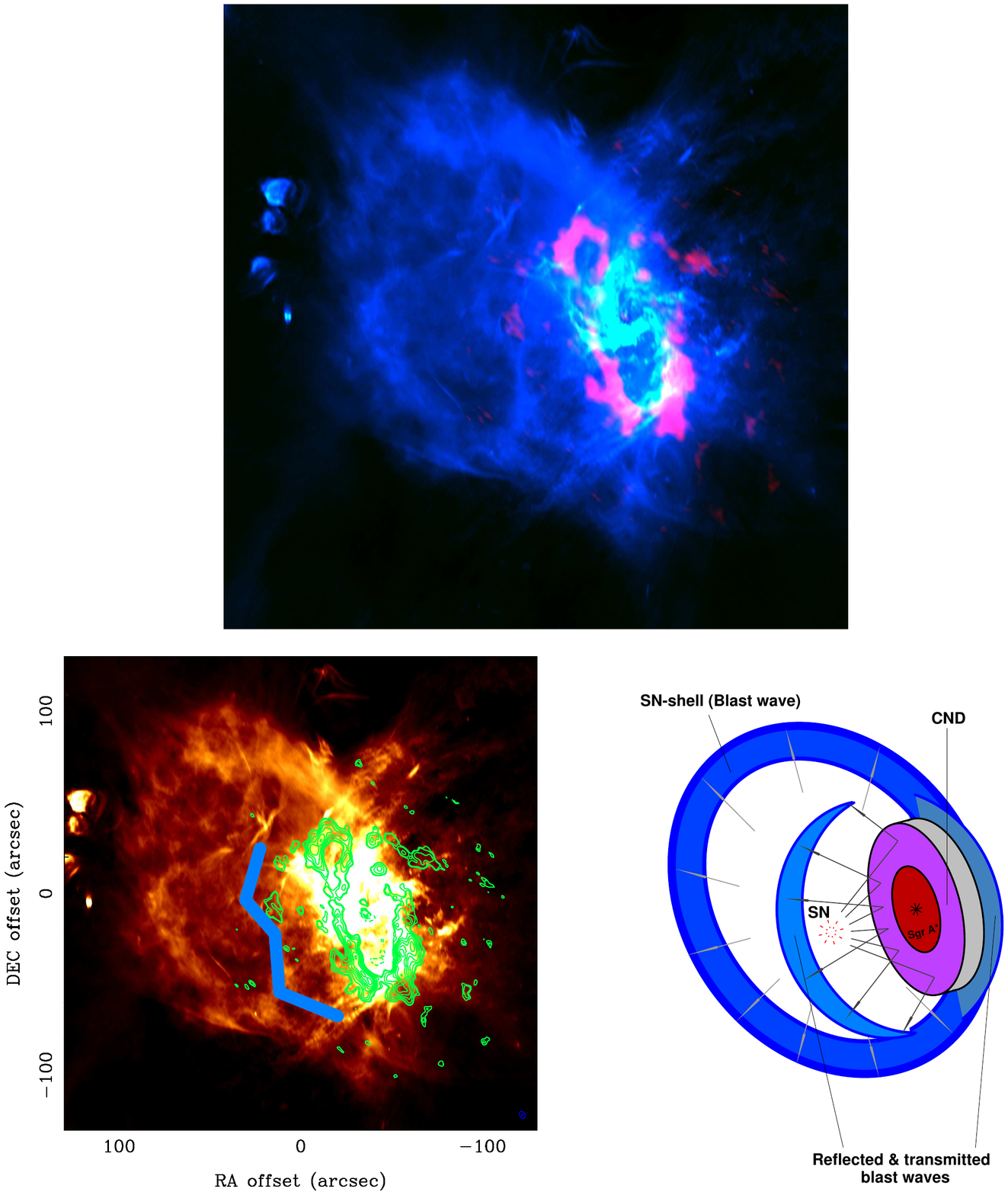}
\caption{{\it Top:} Superposition of the 5.5 GHz JVLA continuum image 
(blue), and the CN molecular line image (red) observed with the 
SMA \citep{mart12}. The combination results in a pink color for the CND. 
Radio continuum emission 
from the Mini-spiral HII gas appears in light blue. {\it Bottom left:} 
The Sgr A East shell and the Sigma Front (marked with four blue line 
segments), with CN emission shown in contours. {\it Bottom right:} 
sketch of our geometrical model illustrating the interaction between 
the SN blast wave and the CND to create noticeable arcs of 
reflected/transmitted blast waves. For the CND, an inclination angle 
of $i=61\arcdeg$ and a major axis position angle of $\phi=19\arcdeg$ 
are assumed. A position angle of 35\arcdeg\ is adopted for the 
major axis of the SNR.
}
\end{figure*}

\begin{figure}[t]
\vskip -5mm
\hskip -25mm
\includegraphics[angle=270,width=135mm]{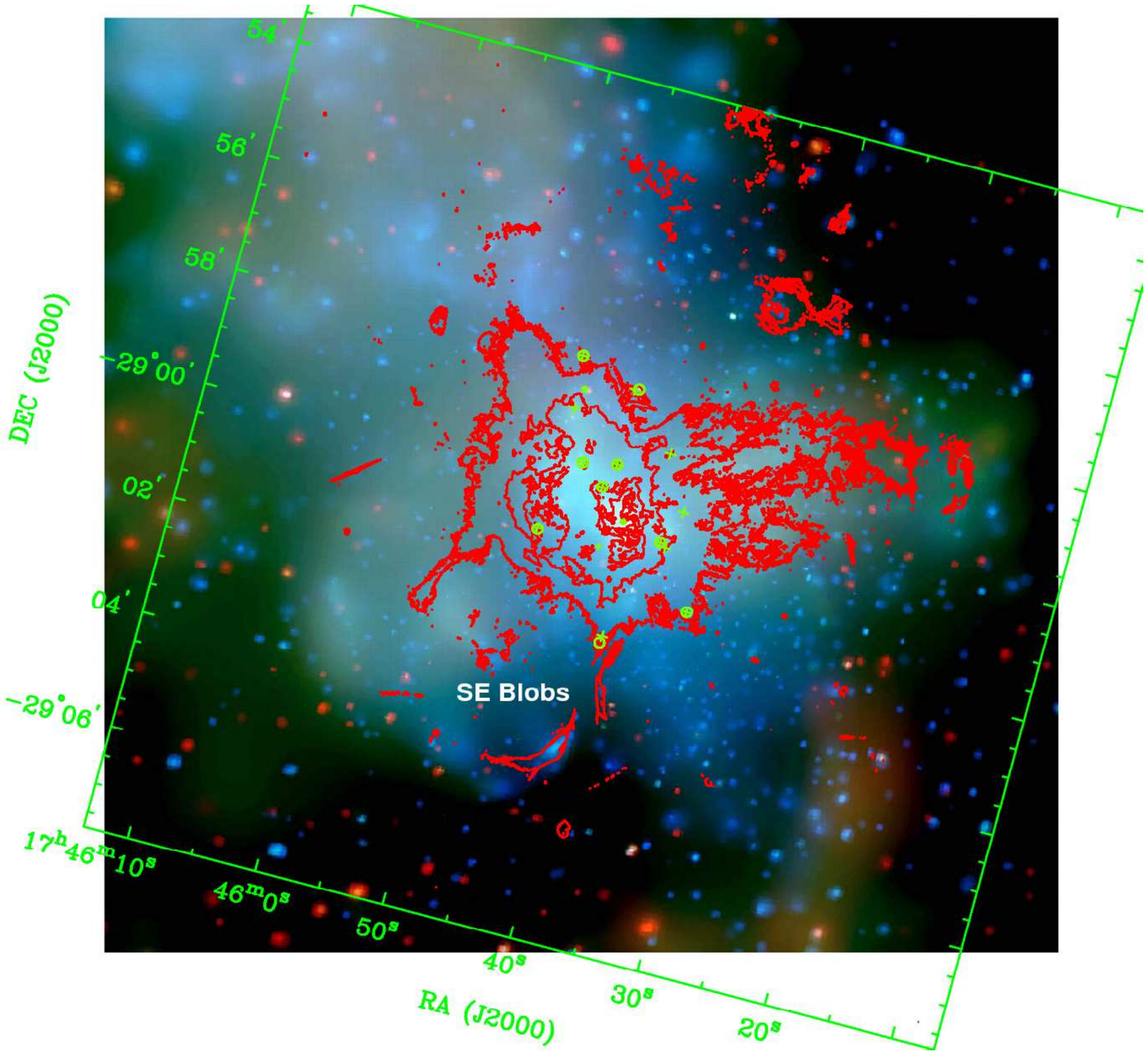}
\vskip -10pt
\caption{Radio contours (from JVLA) overlaid on the Chandra X-ray image
integrated over 1 Msec \citep{mark10}. The coordinate frames between 
the X-ray and radio images were aligned using a dozen X-ray compact 
sources with positions given by \cite{muno09} include Sgr A* and the 
Cannonball \citep{park05,zhao13}.  In this figure, the X-ray and radio
lobes are oriented almost horizontally, or at a position angle of $\sim$~-75
degrees with respect to equatorial north.}
\end{figure}

\begin{figure*}[t]
\vskip -12.5mm
\hskip -15mm
\includegraphics[angle=0,width=210mm]{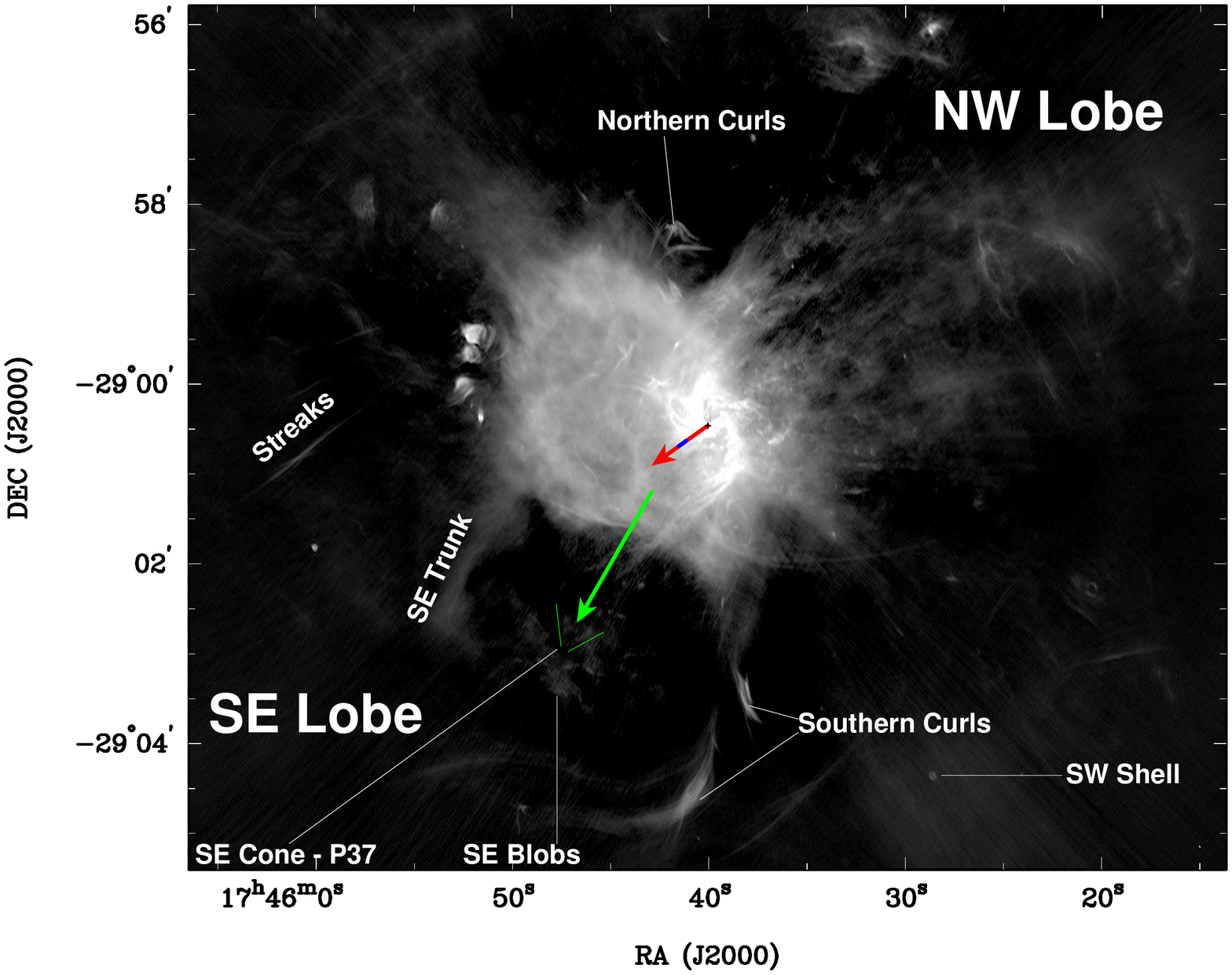} \\
\vskip -15.0mm
\caption{Radio lobes displayed in the JVLA 5.5 GHz image; the red vector 
line indicates the direction of an X-ray feature (blue segment) that has 
been proposed to be part of a jet from Sgr A* \citep{li13}; the green 
vector line points to the tip of the ``V''-shaped feature,
SEblb-5, in the SE Lobe (c.f., section 3.4.2 and Fig. 8.)}
\end{figure*}
 
\subsection{Sigma Front -- a rebound shock from the CND?}
Fig.~3 shows a detailed comparison between our 6 cm continuum images and 
CN line emission from the CND observed by \cite{mart12} with the 
Submillimeter Array (SMA). The CND surrounds the Sgr A West minispiral; 
the shell of the mixed-morphology supernova remnant, Sgr A East, 
appears to encircle both the CND and Sgr A West, although the western 
edge of Sgr A East is ill-defined behind the bright emission from 
Sgr A West. We fit an ellipse to the 150$\sigma$ ($\sigma=8\mu$Jy) 
contour around the Sgr A East shell. The best fit yields a size of 
3.3\arcmin $\times$ 2.3\arcmin\, (7.8pc $\times$ 5.5pc) (P.A.=35\arcdeg) 
and places the apparent center at an offset of 
$\left[\Delta\alpha,~\Delta\delta\right]$ = 
$\left[49.4\arcsec, 19.5\arcsec\right]$ from Sgr A*. Our results appear 
to be in good agreement with an earlier Sgr A East size determination of  
$3.3\arcmin\times2.1\arcmin$ (P.A.$\sim$40\arcdeg)~from lower resolution 
data \citep{pedl89}.

That Sgr A East lies behind Sgr A West has been well established
\citep{yusf87a,pedl89}, but there has been considerable discussion about 
their line-of-sight separation.  A few groups have offered evidence that 
the Sgr A East blast wave has already passed over at least part of Sgr A 
West \citep{herr05, rock05}, and \cite{morr96} had previously pointed 
out that the large-scale asymmetry of the CND could perhaps be attributed 
to the impact of the blast wave. The northeastern portion of the CND, 
which is closest to the centroid of Sgr A East, is truncated relative 
to the near-side southwestern portion.  

Our images reveal another feature that could have resulted from the 
interaction of Sgr A East with the CND.  Located $\sim$1\arcmin\/ 
southeast of Sgr A* or $\sim$0.6\arcmin\/ beyond the southeast 
boundary of the CND, there appears a $\sum-{\rm shaped}$ emission 
feature crossing the middle of the Sgr A East shell at a mean 
$\alpha_{\rm J2000} =17^h45^m43^s.8$, with P.A. $\sim20$\arcdeg 
(Fig.\ 3). This radio feature, to which we hereafter refer as the 
``{\bf Sigma Front}'', can be characterized as four linear 
segments, each about $\sim$ 0.5\arcmin\/ in length with a
typical surface brightness of $\sim$1.5 mJy beam$^{-1}$. 
As summarized in Table 2, the total length and total flux density of 
the Sigma Front are 2\arcmin\ and 0.3 Jy, respectively. We also point 
out that the two ``$\langle$''-shaped structures of the 
Sigma Front bear some correspondence to the shape of the SE edge of the CND. 
More precisely, they are concave toward the most prominent CN-emitting clumps 
located northeast and south of Sgr A* (see Fig.~3). The Sigma Front is also 
clearly present in earlier VLA images obtained at 90, 20 and 6 cm with 
lower angular resolution, see Figs. 3, 4, 5 and 6 of \cite{pedl89}. 
Those authors also notice that this feature is visible on 90 
and 20 cm crosscuts (their Fig. 7).

\subsection{Bipolar Structure}

Fig. 4 shows the radio contours overlaid on the Chandra X-ray image 
integrated over 1 Msec \citep{mark10}, illustrating the bipolar X-ray lobes 
found in the Chandra broad-band X-ray image (3.3-4.7 KeV) by \cite{bag03} 
and \cite{morr03}. The latter authors point out that the soft X-ray bipolar 
lobes are best seen in a map showing the ratio of soft to hard X-rays and 
remark that the shape of the diffuse synchrotron emission observed with 
the VLA at 90 cm \citep{pedl89,nord04} matches the X-ray structure quite 
well. Here, we compare the deep 5.5-GHz JVLA image with the Chandra image 
and with HST/NICMOS imaging of the Paschen-$\alpha$ emission to investigate 
the physical characteristics and the origin of the bipolar lobes. We note 
that the 5.5 GHz JVLA image presented here is best suited for showing 
the fine structure of radio emission while the diffuse emission structure 
as observed on larger scales at 90 cm ($>6.5$\arcmin) is not well sampled 
in the current C+B array observations. 

\begin{table}[b]
\tablenum{2}
\setlength{\tabcolsep}{1.0mm}
\caption{Radio properties of the sources}
\begin{tabular}{lcccc}
\hline\hline \\
{Comp.}&
{$S_{t}$}&
{$S_{p}$}&
{$\Delta\alpha$,$\Delta\delta$} &
{$\Theta_{ma}$,$\Theta_{mi}$,PA}\\
(1)& (2)& (3)& (4)& (5)\\
\hline
\multicolumn{5}{l}{\underbar{SE Wing}} \\
SEwb-1 &\dots&7.7$\pm$1.0&--37.5,--46.0&25,2, --62 \\
SEwb-2 &\dots&6.9$\pm$0.9&--43.5,--48.2&45,2, --56 \\
SEwb-3 &\dots&5.0$\pm$0.7&--51.1,--46.9&56,2, --56 \\
SEwb-4 &\dots&3.3$\pm$0.4&--43.9,--56.3&124,2, --60 \\
\\
\multicolumn{5}{l}{\underbar{NW Wing}} \\
NWwb-1 &\dots&3.5$\pm$0.6&--3,20 &171,2,--50 \\
NWwb-2 &\dots&3.2$\pm$0.5&--20,20 &45,2,--35 \\
NWwb-3 &\dots&9.6$\pm$1.5&--31,20 &129,2,--57 \\
NWwb-4 &\dots&8.5$\pm$1.4&--35,20 &48,2,--59 \\
\\
\multicolumn{5}{l}{\underbar{Sigma Front}} \\
Sf-1 &0.09$\pm$0.01&0.94$\pm$0.15&26.3, 10.6&30,6, --16\\
Sf-2 &0.09$\pm$0.01&1.51$\pm$0.22&22.3,--12.9&25,6, 41\\
Sf-3 &0.16$\pm$0.02&1.43$\pm$0.31&12.9,--38.3&32,6, 5\\
Sf-4 &0.11$\pm$0.01&1.36$\pm$0.27&--4.3,--60.8&35,6, 68\\
Overall& 0.45$\pm$0.06&1.51$\pm$0.22&22.3,--12.9&122,6,25\\
\\
\multicolumn{5}{l}{\underbar{Amorphous radio blobs \& SE Trunk in SE Lobe}} \\
SEblb-1   &0.037$\pm$0.006&0.6$\pm$0.1&59.9,--189.9&10,10, 0\\
SEblb-2   &0.048$\pm$0.012&0.4$\pm$0.1&77.3,--166.5&16,9, 90\\
SEblb-3   &0.060$\pm$0.012&0.5$\pm$0.1&42.9,--174.3&15,10, 90\\
SEblb-4   &0.010$\pm$0.004&0.4$\pm$0.1&17.3,--163.1&14,11, 90\\
SEblb-5   &0.064$\pm$0.013&0.4$\pm$0.1&44.3,--147.5&15,15, 0\\
SEblb-6   &0.033$\pm$0.005&0.7$\pm$0.1&30.6,--134.8&16,11, 90\\
SEblb-7   &0.036$\pm$0.007&0.4$\pm$0.1&38.0,--104.1&12.5,9, 90\\
SEblb-8   &0.02$\pm$0.004&0.5$\pm$0.1&9.5,--116.7&9,5.5, 90\\
Overall& 1.1$\pm$0.2   &0.7$\pm$0.1&30.6,--134.8&150,120,150\\
SEtr   &0.54$\pm$0.05 &0.5$\pm$0.1&136,--132&115,15,152\\
\\
\multicolumn{5}{l}{\underbar{Smoke Rings in NW Lobe}} \\
NWstr-1a &3.1$\pm$0.3&3.6$\pm$0.5& --62.8,67.8 & 160,56,--58  \\
NWstr-1b &0.51$\pm$0.08&0.7$\pm$0.1& --200.7,124.4& 86,22,-100  \\
NWstr-2 &3.7$\pm$0.4&3.1$\pm$0.3&--69.6,16.1&340,50,--67  \\
NWstr-3 &2.5$\pm$0.2&10.0$\pm$0.5&--76.2,-21.5&210,58,--78\\
SmR-1  &0.93$\pm$0.09&1.8$\pm$0.2&--353.9,92.7&80,30,18 \\
SmR-2  &1.0$\pm$0.1&1.8$\pm$0.2&-283.5,107.4&80,30,55 \\
SmR-3  &0.41$\pm$0.06&1.4$\pm$0.2&-219.3,104.4&60,26,28  \\
Overall&17$\pm$3&10.0$\pm$0.5&--76.2,-21.5&380,190,--70\\
\hline
\end{tabular}\\
\begin{tabular}{p{0.45\textwidth}}
{\footnotesize Column 1 is the name of component sources;
and Columns 2 -- 5 respectively list their radio properties: flux 
density (Jy), peak intensity (mJy beam$^{-1}$), position 
offsets (arcsec) from the field center (given in the 
footnote of Table 1), full width (arcsec) of a source along 
the maximum and minimum extents, and position angle
of the longest extent (degrees); for a compact source.
The flux density for each of the emission components is 
determined by integrating the emission intensity over
the area given in Column 5 with a correction for background 
and/or foreground contaminations. The correction utilizes the 
mean intensity estimated from a nearby off-source area 
corresponding to each source. 
}\\
\end{tabular}
\end{table}

\subsubsection{NW Radio Lobe}

The 5.5 GHz image (Fig. 4) shows that the NW radio lobe is closely 
associated with the NW X-ray lobe that extends to positive latitudes, 
with a length in the radio of 360\arcsec~(14 pc) and a maximum width 
of 180\arcsec~(7 pc) at the base across Sgr A West. The NW radio lobe 
contains several elongated continuum emission streamers at distances 
from 100\arcsec\ to 200\arcsec\ from Sgr A* \citep{yusf87a}, and bright 
ring-like emission nebulae -- ``{\bf Smoke Rings}'' -- located at 
greater distances (Figs.\ 5 \& 6).  All but one of the NW streamers 
are generally perpendicular to the Galactic plane, PA=27\arcdeg, 
\citep{binn98}, although they show some curvature. Four of these 
features are identified in Fig.\ 6. The northernmost streamer appears 
to consist of two segments; NWstr-1a appears to be a continuation of 
the NW Wing, with strong continuum emission (3.1$\pm$0.3 Jy) arising 
from a 2.7\arcmin$\times$1\arcmin\/ region oriented in PA --58\arcdeg; 
this streamer apparently curves over to a weaker (0.5$\pm$0.1 Jy) and 
thinner segment, NWstr-1b, of size 1.5\arcmin$\times$0.4\arcmin, and 
PA $\sim$ --100\arcdeg.  The radio emission of NWstr-1a appears to be 
peaked at a short filament (NW Streak) in Fig. 2 of \cite{zhao13} 
(marked with open arrow in Fig. 6 of this paper), which is associated 
with the compact X-ray source, CXOUGCJ174538.0-285911 
\citep{muno09}. 

The long, straight streamer NWstr-2 runs through the middle of the NW Lobe
with a relative constant position angle of ~(--67\arcdeg). NWstr-2 is 
therefore oriented within 4\arcdeg\ of the Galaxy's rotation axis, and 
is almost exactly co-aligned with Sgr A*.  The flux density integrated 
over its 5.7\arcmin$\times$0.8\arcmin\/ size is 4.7$\pm$0.4 Jy. The 
highest-latitude extensions of NWstr-2 appear to intersect, and perhaps 
produce, the Smoke Rings. 

The southernmost and least well-defined streamer is NWstr-3. The flux 
density is 2.5$\pm$0.2 Jy integrated within a region of size 
3.5\arcmin$\times$1\arcmin, PA $\sim$ --78\arcdeg.  Near the base of 
NWstr-3, there is a bright head-tail source, source M, which might be 
physically related to NWstr-3.  Source M was identified as a thermal 
emission arch based on observations of both the $\left[\rm NeII\right]$ 
line \citep{sera84} and radio continuum \citep{yusf87a}. The compact 
emission in source M is identified with source 2 in Fig. 2 of \cite{zhao13}, 
which appears to be associated with the compact X-ray source 
CXOUGCJ174536.9-290039 \citep{muno09}.  The radio intensity of the 
unresolved head of source M is 10$\pm0.5$ mJy beam$^{-1}$. The tail 
extends toward positive latitudes, but the orientation is misaligned 
with NWstr-3 by $\sim$20\arcdeg. We will present more detailed 
observations of source M in a subsequent publication.   

Near the base of the NW Lobe, at distances from Sgr A* of 50\arcsec\ -- 
180\arcsec\ (2 -- 7.5 pc), there is a field of amorphous radio blobs, 
or {\bf NW blobs}, some of which have a bowshock morphology suggestive 
of a shock interaction between a wind and ambient gas clumps. The 
curvature of several of these apparent shocks is convex toward Sgr A*, 
indicating that the wind emanates from Sgr A* itself, from the central 
cluster of massive young stars, or from some combination of the two. 
A similar field of amorphous blobs occurs in the SE Lobe, described below. 
Such blobs, with typical surface brightnesses $<$0.5 mJy beam$^{-1}$,
are largely concentrated in the radio lobes, and there are far fewer 
instances of them elsewhere around Sgr A. This distribution is consistent 
with the hypothesis that the lobes have been sculpted by outflowing 
winds that are interacting with ambient density enhancements. Further 
investigation with shock diagnostics is needed to address the following 
questions concerning the origins of the radio blobs and streamers: 
is the emission from the amorphous blobs indeed a result of strong 
shocks; or alternatively, is the emission dominated by free-free radiation 
from gas clumps that have been ionized by the UV radiation from the 
massive young star cluster in the central parsec. In a subsequent paper, 
we will present a closer examination of the amorphous radio blobs 
in the lobes.  

Most of the radio features within the lobes appear to have closely matching 
Paschen-$\alpha$ (hereafter, P$\alpha$) counterparts, indicating that their 
emission is thermal free-free radiation. This correlation is broadly 
illustrated by the bottom-right inset of Fig. 6, which shows the HST/NICMOS 
image of P$\alpha$ emission \citep{wang10,dong11} overlaid with radio contours 
covering the region containing the NW radio lobe and the localized emission 
clumps in both the NW and SE lobes.  Both the P$\alpha$ and radio images in 
Fig. 6 are smoothed to a circular beam of 2\arcsec. The close 
correspondence of the radio and P$\alpha$ in the outer portions of the
NW Lobe is shown in detail in Fig. 7, which shows a succession of 
three nebulae, spaced along the central axis of the NW Lobe.  
In particular, the radio morphology of these three nebulae 
(``Smoke Rings'', see top image in Fig. 7), matches well with 
that observed in P$\alpha$ (Fig. 7, bottom) when the patchy 
foreground extinction is taken into account. 

The integrated flux densities of SmR-1, SmR-2 and SmR-3, respectively 
in order of decreasing distance from Sgr A*, are 0.93$\pm$0.09, 
1.0$\pm$0.1 and 0.41$\pm$0.06 Jy. The extent of these nebulae along 
their major axes are in the range 1\arcmin~to 1.3\arcmin.  
The major axes of SmR-1 and SmR-3 appear to be nearly perpendicular to
the central axis of the NW radio lobe while the major axis of SmR-2 
appears to be tilted by $\sim$50\arcdeg\/ with respect to the 
central axis of the NW radio lobe. In addition, a few thin radio 
filaments can be matched with P$\alpha$ filaments, indicating
that not all of the radio filaments identified in this region 
are nonthermal emitters \citep{mzg14}. Two high-mass, X-ray-emitting 
emission-line stars (green circles in Fig. 7) have been identified 
in the vicinity of the Smoke Rings \citep{mau10,dong12}, one located 
at the SE rim of SmR-1 (CXOGC J174516.7-285824, or P134) and the other 
located between SmR-2 and SmR-3 (CXOGC J174522.6-285844, or P39). 
These stars possibly contribute to the ionization of the Smoke Ring 
nebulae and perhaps the NW streamers. However, the NW and SW lobes 
have relatively soft, thermal X-ray counterparts (\cite{morr03}; 
\cite{pont15}, and references therein; see Figs.\ 18 and 19 of the 
latter reference), indicating that at least some of the plasma 
in these nebulae is far hotter than expected for an HII region 
ionized only by hot stars. Consequently, strong shocks produced 
by high-velocity impacts of winds from the central region on 
ambient density enhancements seem to be a more likely mechanism
for producing both the ionization and the X-ray emission.

The overall radio flux density from the NW Lobe is 17$\pm$3 Jy in 
a region of size 6.3\arcmin$\times$3.2\arcmin\/(15 pc $\times$ 7.4 pc), 
with the long dimension at a position angle of --70\arcdeg, and with 
the bottom of this rectangle located 0.3\arcmin\/ toward positive 
Galactic latitudes from Sgr A*.  

\subsubsection{SE Radio Lobe}

The radio counterpart of the SE X-ray lobe is much less prominent 
than that of the NW Lobe; indeed, without the clear presence of 
the SE {\it X-ray} lobe to help define it \citep{pont15}, the physical 
coherence of the SE {\it radio} lobe is questionable. The X-ray 
emission from the SE Lobe appears to be outlined by the radio filaments 
labelled in Fig.\ 5 -- the ``Streaks'' (a group of filaments) and 
the ``Southern Curls'' \citep{mzg14}, although this could be a chance 
superposition. Also, a ridge of emission (labelled {\bf SE Trunk} 
in Fig.\ 5) cuts across the SE radio lobe and might be part of this 
structure, although its nature and origin are unclear. The amorphous 
radio blobs in the SE Lobe located between the SE Trunk 
and the filamentary Southern Curls show significant 
P$\alpha$, as illustrated in Figs. 8a (radio) and 8b (P$\alpha$).  
Here, the IR image has been smoothed to the same resolution as the 
radio of 2\arcsec.

The inset to Fig. 8b (top right) shows a detail of a presumably foreground 
shell source, the ``SW Shell'', located to the southwest of Sgr A 
in our 5.5 GHz image and marked in Fig. 5. The close correspondence 
of the radio (color) with the P$\alpha$ (contours) shows that the 
two coordinate frames are well aligned.

The eight panels at the bottom of Fig. 8 show the radio emission 
(color) from eight of the amorphous radio blobs, the ``{\bf SE Blobs}'', 
overlaid with P$\alpha$ emission contours. SEblb-1 corresponds to 
the region surrounding the position of a bright star present in the 
1.87 $\mu$m images (but imperfectly removed in the continuum-subtracted 
image, Fig. 8b). A small arc of radio emission curves around
the location of the star toward the NW, suggesting that SEblb-1 is locally 
produced by winds and/or radiation from this star.

The conical shape of SEblb-5 is present in both radio and P$\alpha$, 
suggesting the presence of a bow shock. The apex of this apparent cone 
(SE Cone) points away from Sgr A*, which raises the possibility that 
it has resulted from a collimated outflow from Sgr A*. However, the 
direction toward which the apex is pointing is offset by about 25\arcdeg\ 
from that of the jet proposed by \citep{li13} (see Fig. 5);
no outflows have been observed or proposed in this direction. 

A filamentary radio feature in SEblb-2 appears to be closely coincident 
with the P$\alpha$ emission structure. On the other hand, SEblb-3 shows 
another filamentary radio structure, with no discernible P$\alpha$ 
counterpart. The thin, curved radio filament of SEblb-4 also appears to 
be closely matched by the P$\alpha$ emission over at least part of its 
length. SEblb-6 shows a complex radio morphology surrounding a compact 
source, and diffuse P$\alpha$ emission permeates the entire structure. 
SEblb-7 has filamentary emission in both radio and P$\alpha$, 
but it is much more continuous in the radio. SEblb-8 consists of 
two clumps in the radio; only the western component has a 
P$\alpha$ counterpart. The radio properties of the amorphous blobs 
are summarized in Table 2. The overall flux density of the blobs in 
the southern lobe is 1.1$\pm$0.2 Jy, within a region of angular 
size 2.5\arcmin$\times$2\arcmin.

The SE trunk (SEtr), located toward the NE of the SE Blobs (Fig.\ 5), 
subtends 115\arcsec$\times$15\arcsec~at a position angle of $\sim$152\arcdeg) 
and has a total 5.5 GHz flux density of 0.54$\pm$0.05 Jy. This feature 
shows no significant counterpart in P$\alpha$ emission, so it might be a 
nonthermally emitting structure. Indeed, the SEtr was also identified in 
the 90 and 20 cm images of \cite{pedl89} as a $\sim$1\arcmin-scale structure 
at $\alpha_{\rm B1950}=17^h42^m40^s$, $\delta_{\rm B1950}=-29\arcdeg01\arcmin$ 
in the Sgr A halo. At those wavelengths, the SEtr appears to have about the same
spectral index as the supernova remnant, Sgr A East.  

\begin{figure*}[t]
\centering
\includegraphics[angle=0,width=180mm]{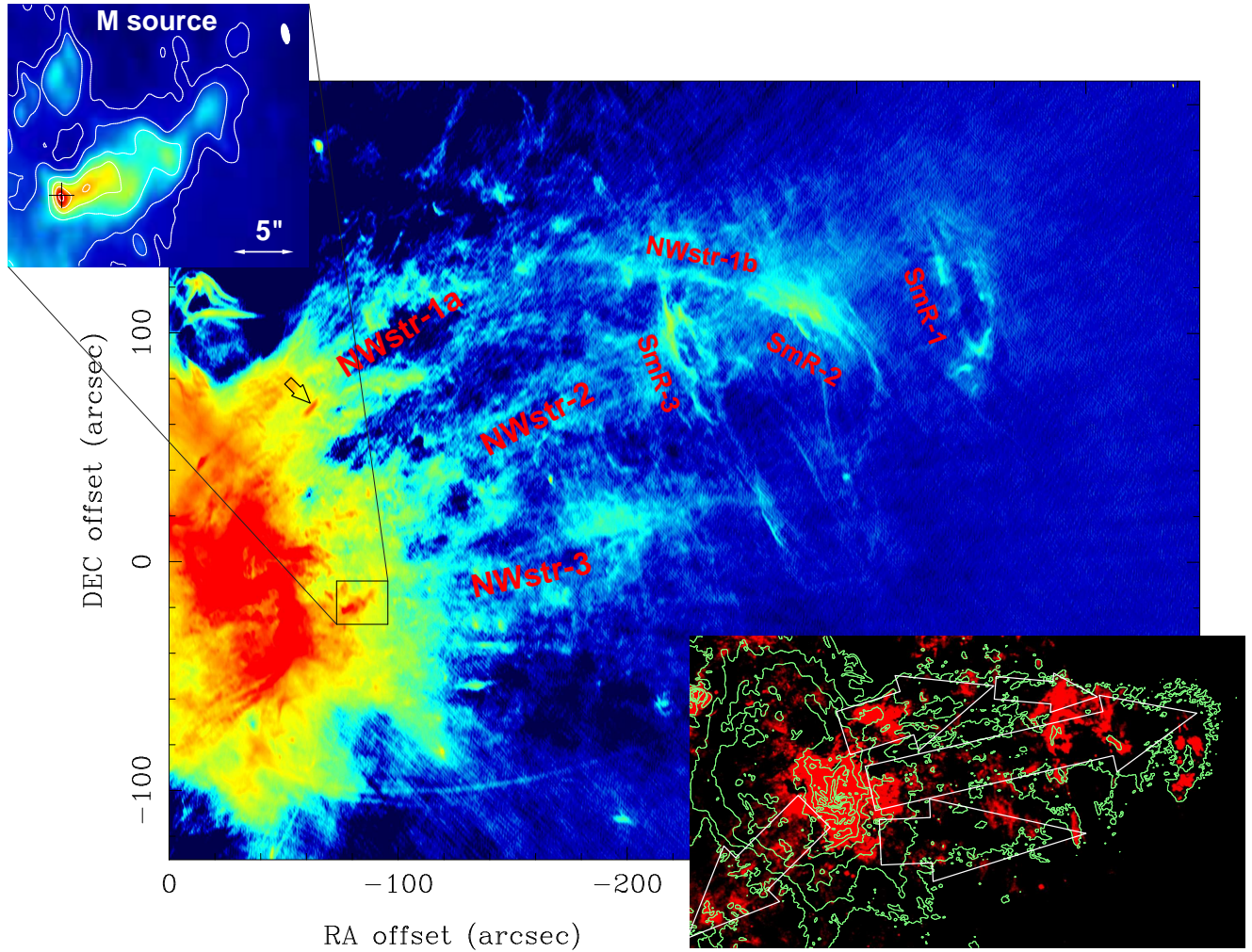} \\
\vskip 0pt
\caption{The 5.5 GHz radio image showing the emission streamers
as labelled in the NW radio lobe with the FWHM beam of 
1.6\arcsec$\times$0.6\arcsec\ (PA=11\arcdeg).  Also the radio 
counterpart of the compact X-ray source CXOUGCJ174538.0-285911 
is marked with an open arrow. The inset at top left shows the 
head-tail structure of the southern arch of source "M", located 
in the region marked with open rectangle. The contours of this inset 
are 1$\sigma\times$(100, 150, 250, 400, 600, 850), $\sigma=8\mu$ Jy 
beam$^{-1}$, and the beam is marked on the top-right corner. The 
bottom-right inset presents the radio image as contours (at 10$\mu$Jy 
beam$^{-1}\times$ 2$^n$ and $n=2, 7, 9, 10, 11, ..., 16$) overlaid on 
the HST/NICMOS P$\alpha$ image (color) \citep{wang10}. Both  radio and 
P$\alpha$ images have been smoothed with a circular Gaussian of 
FWHM 2\arcsec. The equatorial coordinate frame in this inset has been 
rotated 15\arcdeg (clockwise). The directions of the major axes of 
the radio streamers in the NW Lobe as well as the predominant axis 
of the amorphous blobs in the SW Lobe are marked by open arrows.}
\end{figure*}

\begin{figure}[t]
\vskip -5mm
\hskip -7.5mm
\includegraphics[angle=0,width=97.5mm]{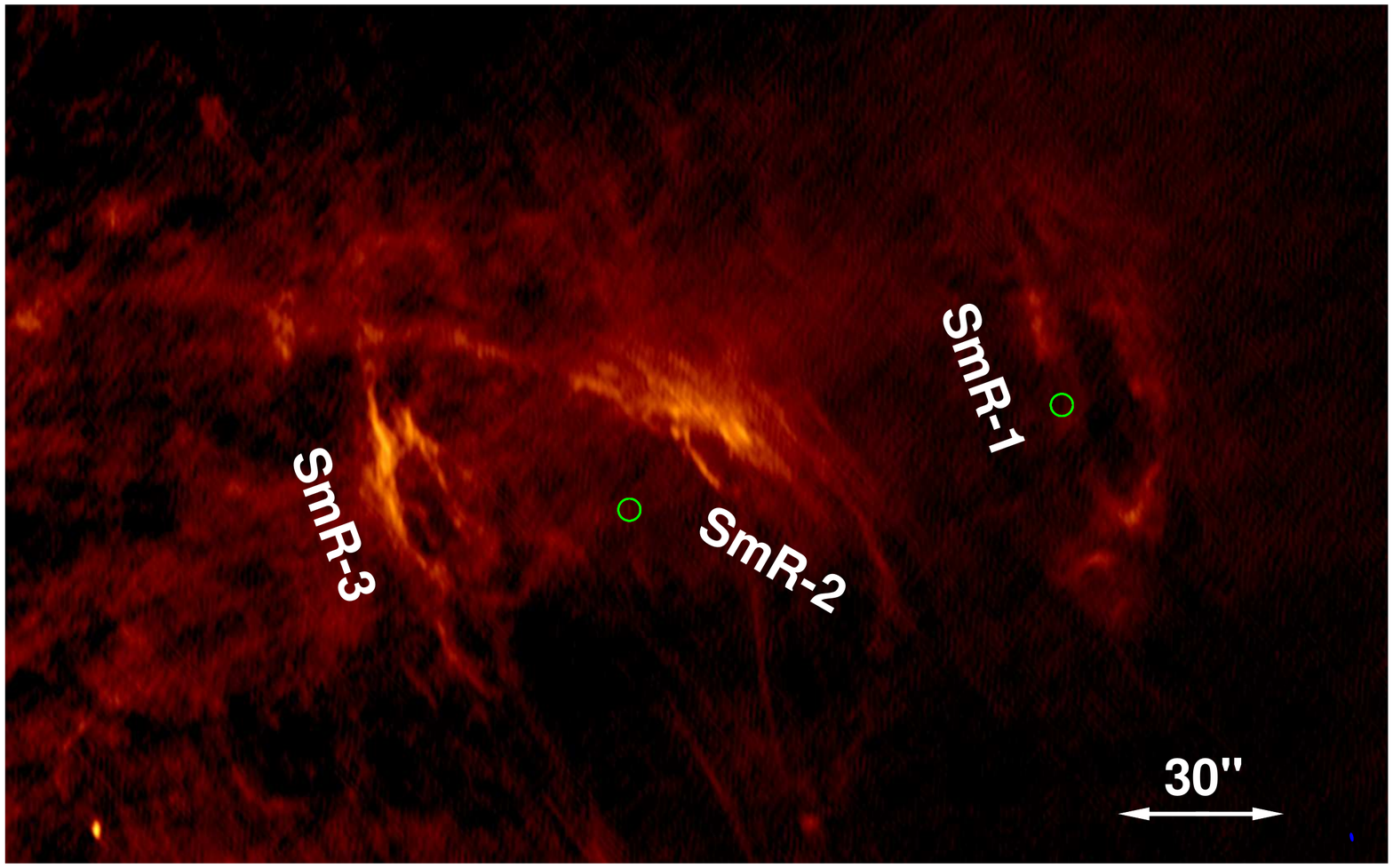} \\
\vskip -25mm
\hskip -7.5mm
\includegraphics[angle=0,width=97.5mm]{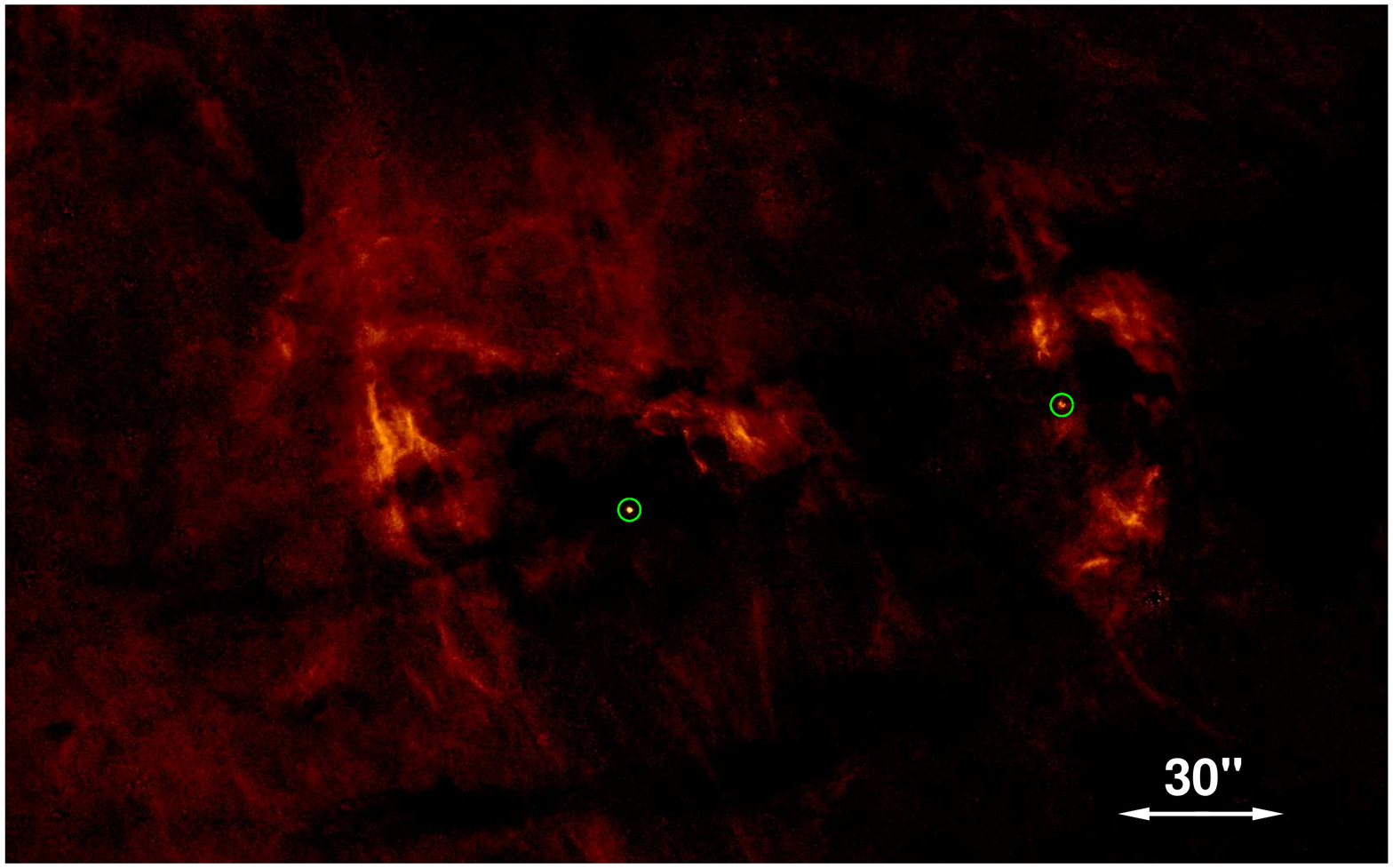}
\vskip -12.5mm 
\caption{
Top: VLA image of 5.5 GHz radio emission from the outermost portions of 
the NW Lobe, showing details of the structures that we refer to as 
"Smoke Rings" (SmR 1, SmR 2 and SmR 3, from right to left).  Equatorial 
North is up. The FWHM beam is
1.6\arcsec$\times$0.6\arcsec\ (PA=11\arcdeg). 
No correction for primary beam attenuation has been applied.
Bottom: The HST/NICMOS image of P$\alpha$ emission \citep{wang10,dong11}
corresponding to the same region. Minor differences between the two 
images can be attributed to foreground extinction features that affect 
the P$\alpha$ distribution. The green circles mark two massive stars in 
the region, as discussed in the text. 
}
\end{figure}

\begin{figure*}[t]
\vskip -5mm
\includegraphics[angle=0,width=87.5mm]{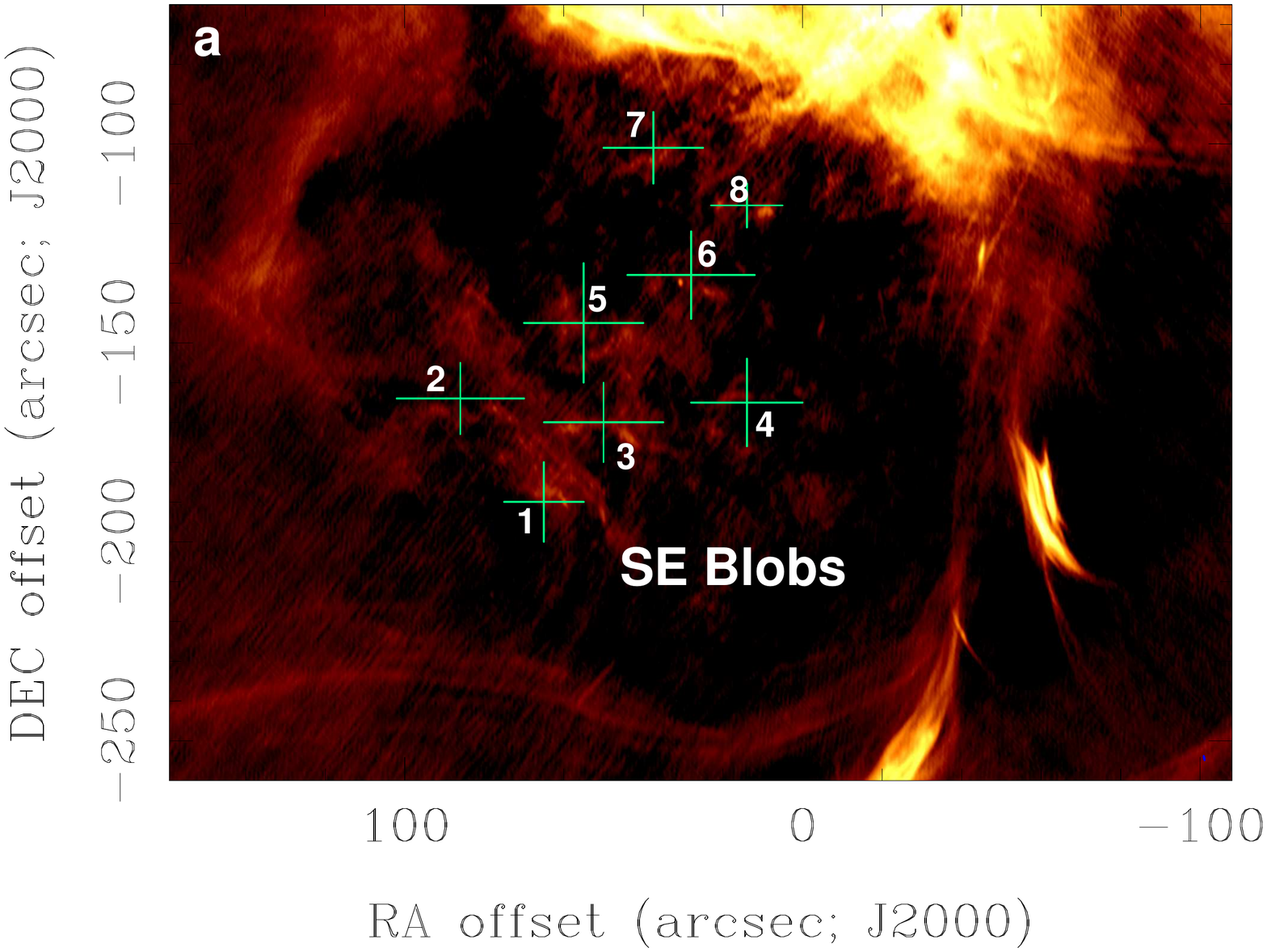}
\includegraphics[angle=0,width=87.5mm]{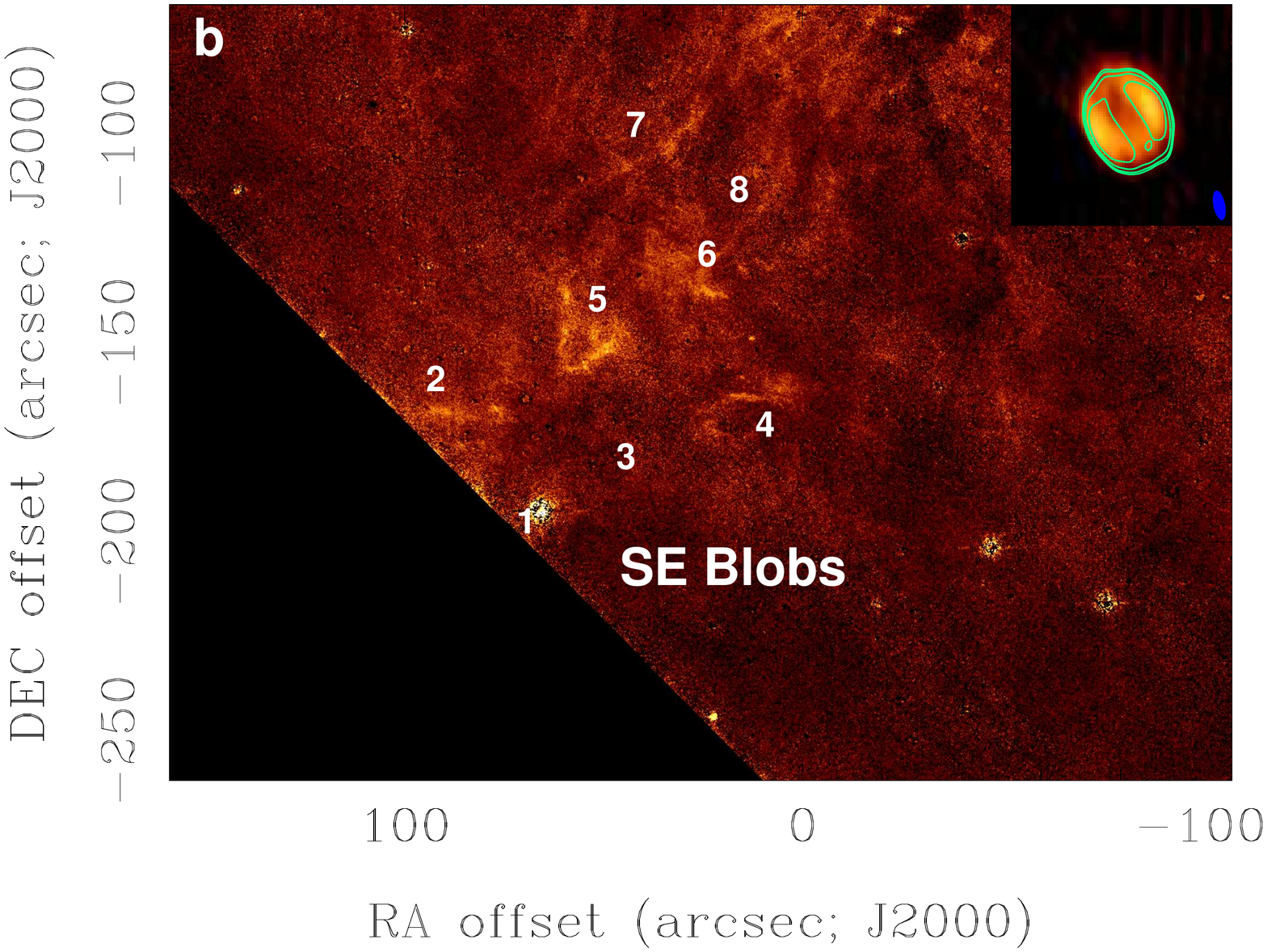}\\
\vskip -47.5mm
\hskip -5mm
\includegraphics[angle=-90,width=178mm]{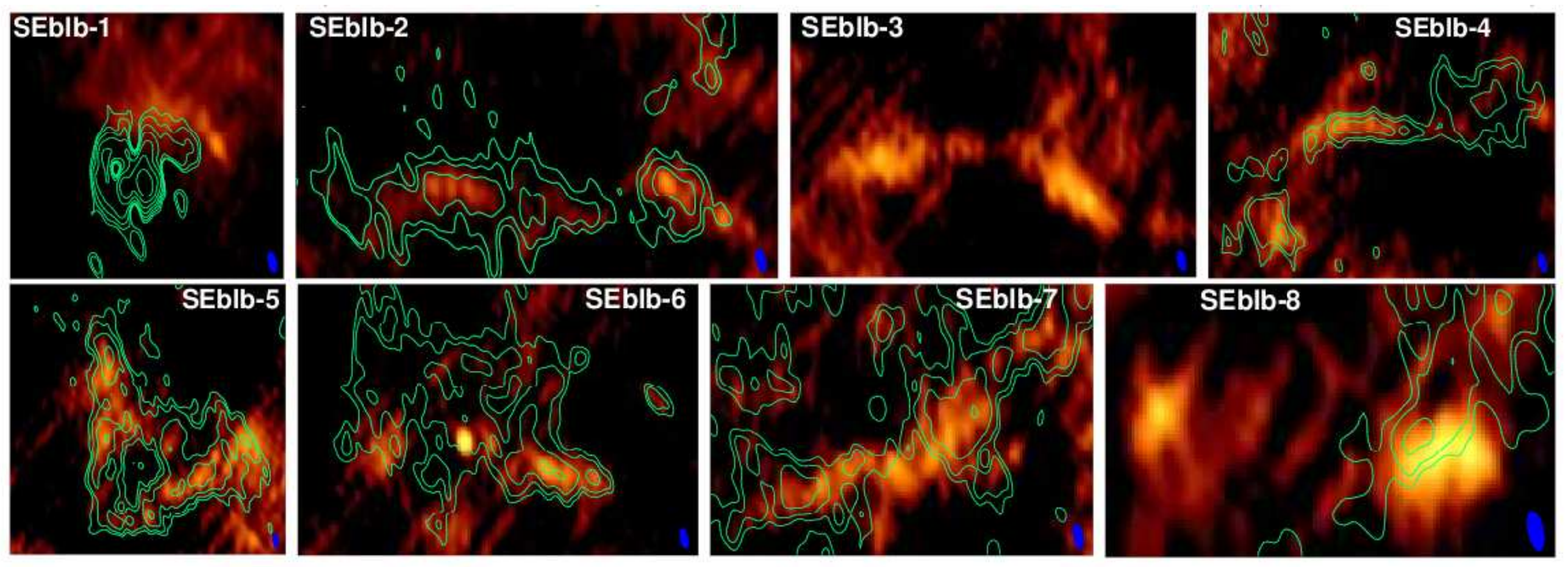}
\vskip -40mm 
\caption{Top panel $a$:  a section of the radio image showing
the distribution of amorphous radio blobs in the SE Lobe. 
The FWHM beam of
1.6\arcsec$\times$0.6\arcsec\ (PA=11\arcdeg).
Top panel 
$b$: an HST/NICMOS image showing the P$\alpha$ counterpart of the 
same region. The individual components are marked with their sequential 
numbers in both images. The inset at top-right shows an image of the 
SW Shell (marked in Fig. 5) in radio (color) and P$\alpha$ (contours) 
to verify the alignment between the radio and IR coordinate frames.  
The bottom eight panels show the details of individual components
of the amorphous radio blobs (color) overlaid with contours of
P$\alpha$ emission that has been smoothed with a 2-D Gaussian
function identical to the radio clean beam of 
1.6\arcsec$\times$0.6\arcsec\ (PA=11\arcdeg). The central positions
(offsets in arcseconds from the field center, see footnote of Table 1) of
the eight regions are (65,--190), (86,--164), (50,--170), (14,--165),
(55,--145), (28,--133), (37.5,--101) and (14,--115.5).}
\end{figure*}

\subsection{The Sgr A Halo}
The SE and NW radio lobes lie within the radio halo of Sgr A, previously 
imaged by \cite{yusf87a} at 20 cm and by \cite{pedl89} at 90 cm.  
To interpret the shallow turnover at frequencies below 500 MHz, \cite{pedl89}
argued that nonthermally emitting particles are uniformly mixed with the 
thermal gas, with a total halo mass of $\sim3\times10^3$ M$\odot$.
The diffuse continuum radiation of the halo, with a steep spectral index 
of $\alpha=-0.7$ and a flux density of $\sim$130 Jy at 6 cm, is dominated 
by nonthermal emission \citep{pedl89}. This extended emission is not well 
sampled in the high resolution VLA observations of this paper. Comparison 
of our 5.5 GHz image with the P$\alpha$ image indicates that most of the 
thermal gas appears to be in the amorphous gas clumps in the NW and SE lobes. 
From the measured flux density at 5.5 GHz, the NW Lobe alone appears to 
account for an HII mass of 
$\sim2.6\times10^3~{\rm M_{\odot}}~f_{\rm V}^{1/2}\zeta_{ff}^{1/2}$, where
$f_{\rm V}$ and $\zeta_{ff}$ are, respectively, the volume filling factor 
of free-free emitting gas and the thermal fraction of the total flux 
density at 5.5 GHz (see Appendix B). 

\section{Astrophysical Implications}
\subsection{A Reflected Blast Wave Front in The SNR}
The iron enrichment of Sgr A East inferred from X-ray studies 
\citep{maed02, park05, koya07}, indicates that this nonthermal bubble 
originated as a type II supernova. Sgr A East appears to be a relatively 
old SNR with an age of $\sim$10,000 yrs \citep{maed02,zhao13}, old enough 
that the reverse shock has propagated all the way back to the center of 
the remnant. The recent detection of a warm dust clump in Sgr A East 
has been interpreted as evidence that the dust produced in a SNR can 
survive the passage of the reverse shock \citep{lau15}. However, the 
rebound shock resulting from the presumed recent impact of the blast 
wave on the CND will not have had time to travel back to the center of 
the remnant.  We hypothesize that the Sigma Front is the rebound shock 
resulting from the reflection of the SN blast wave (BW) off the massive 
CND. The CND consists of a relatively dense medium traced by molecular 
species such as HCN \citep{gust87,jack93} and CN \citep{mart12}.  The CND 
forms an annular ring between radii of 1.5 - 2.3 pc \citep{jack93} and 
up to a radius $>$5 pc for low excitation lines \citep{gust87}.

The interaction of a BW and a dense cloud in the ISM has been studied 
over the past four decades, using both theoretically calculations, {\it e.g.}, 
by \cite{mcke75}, \cite{sgro75}, \cite{silk73} and \cite{spit82}, and 
interpretations of SNR observations, {\it e.g.} \cite{miya01} for 
the Cygnus Loop. As a supernova blast wave strikes an interstellar gas 
cloud, two shocks are produced: (1) a reflected shock propagating back 
into the medium of the SNR and (2) a transmitted shock propagating into 
the cloud \citep{sgro75}. In this paper, we investigate this possibility 
with a simple geometrical model for the 
location of the Sgr A East SN with respect to the CND and Sgr A West. 
Based on the geometrical model, we sketch in the bottom-right panel 
of Fig. 3 a cartoon graphic to illustrate the geometrical configuration 
of the reflected/transmitted SN BWs relative to the CND and the SNR shell.
In order to further assess the applicability of this geometrical model
to the Galactic center, the following caveats must be considered:

First, the relative physical placements of Sgr A East and the CND, and of
    the BW and the CND, constrain the location of the SN explosion site to 
    be a projected distance of 2 pc east of the CND, and relatively close 
    in the background \citep{herr05}. But because the Sigma Front is east 
    of the CND indicating that the BW has been reflected back toward the 
    center of the SNR in projection, and because the eastern side of the 
    CND is the far side, the Sigma Front can only be accounted for in this 
    model if the BW has rebounded. Our model assumes that the BW has recently 
    passed over Sgr A* \citep{herr05, rock05}, in which case the maximum 3D 
    separation of the center of the SNR from Sgr A* is $d_{\rm max} =3.9$ pc, 
    which is the semi-major axis of the SNR.  This is consistent with the 
    lower limit of the distance $d=3.6^{+0.7}_{-0.7}$ pc between the SN 
    dust clumps and Sgr A* determined from the best-fit models for the 
    observed dust SED \citep{lau15}, if the $\sim$1 pc projected offset of 
    the dust concentration from the apparent center of the SNR is taken 
    into account. The distance of the SN  behind the sky plane containing 
    Sgr A* is then $\sim \sqrt{d_{\rm max}^2-(2\/{\rm pc})^2} \approx 3.3$ pc.  

Second, the interaction can deform the projected shell
    of the BW front from a circular shape, but the predicted minor 
    to major axis size ratio, or the compression factor, of 0.85 
    based on a simple geometrical model assuming an interaction
    between the BW and the CND appears to be insufficient
    to explain the observed ratio of $\sim0.69$; this could be caused by 
    stretching of the major axis by the strong tidal shear in this region \citep{uchi98}.

Third, the distribution of the dense clumps in the CND may be responsible for
    the observed $\sum$-shape of the Sigma Front. Both the northern and  
    southern ``$\langle$'' features likely correspond to the BWs reflected 
    from the two high-density ($n\left[{\rm H_2}\right]>10^6$ cm$^{-3}$) 
    molecular regions in the CND situated at projected distances of 
    $\sim2$ pc NE and SW of Sgr A*, as shown in the papers by \cite{jack93} 
    and \cite{mart12}; see also the CN intensity image depicted in Fig. 3.
 
Fourth, the inclination ($i\sim50\arcdeg-70\arcdeg$) and major axis position 
    angle ($\phi\sim25\arcdeg$) of the CND were determined by fitting the 
    observed radial velocities with a circular rotation model 
    \citep{gust87,jack93} while mean values of 
    $\bar{i}\sim61\arcdeg$ and $\bar\phi\sim19\arcdeg$ are found by fitting 
    the radial velocities of radio recombination line images with Keplerian 
    orbits around the central black hole \citep{zhao09,zhao10}. The results 
    from molecular line and RRL observations appear to be in good agreement. 
    We adopt $\bar{i}\sim61\arcdeg$ and $\bar\phi=19$\arcdeg~in the simple 
    geometrical model considered here. Thus in the illustration (bottom-right 
    of Fig. 3), a position angle of 19\arcdeg\ is used for the major axis 
    of the CND, and 35\arcdeg\ for that of the SNR shell.

Fifth, this model suggests that the interactions of the  BWs with the CND
    separate the radio emitting gas energized by the BWs into two parts:
    (1) the reflected BW manifested as the Sigma Front, and (2) the 
    transmitted BW lying behind and west (to north) closer to the CND, 
    likely responsible for the emission arc around the Sgr A West absorption 
    feature as observed at 90 cm, best viewed  in color as shown in Fig. 1 
    of \cite{goss89}. The overall structure of the Sgr A West absorption 
    against Sgr A East, as observed at 90 cm with resolutions of 
    12\arcsec~\citep{pedl89,goss89} and 43\arcsec$\times$24\arcsec\/ 
    \citep{laro00}, can be explained by a simple geometrical model in 
    which most of the radiation from the transmitted BW that is located 
    NW of Sgr A* and behind the CND, is absorbed by the thermal gas in 
    Sgr A West located in front of the transmitted BWs.

Finally, we address the question of whether the SE and NW Wings might 
    have been affected by the BW. As described in section 3.2, these 
    features are associated with the ionized gas of the Western Arc and 
    Northern Arm in Sgr A West, respectively. Both the Western Arc and 
    Northern Arm are co-planar with the CND \citep{paum04,zhao09,zhao10}.
    The relatively low-density ionized gas in the SE and NW wings appears 
    to have been expelled, probably by the nuclear wind, from the CND to 
    move outward along bipolar direction (see section 4.2). For a large 
    inclination angle of the CND, the dominant motion of the expelled material 
    would then be in the sky plane. Given $i=60$\arcdeg, then for the observed 
    projected length of the wings, $L_{\rm w, xy}=2$\arcmin\/($\sim$5 pc), 
    the corresponding scale along the line of sight is $L_{\rm w, z}=1.2$\arcmin\/($\sim$3 pc). 
    Thus, the SE Wing would be situated sufficiently in front of the SN shell 
    to not have been disturbed by the BW, while the NW Wing might have already 
    interacted with the transmitted BW. 

\subsection{Ionized Outflow from The CND}

One of the prominent features in the RBZ at the Galactic center is
the bipolar structure. Although the SE Lobe is nearly engulfed in the
Sgr A East SNR, the radio emission from the NW Lobe delineates a collimated
outflow structure that appears to be dominated by free-free emission at 6 cm.
Ionized outflows are an ubiquitous phenomenon in the nuclear regions of 
galaxies, from QSOs \citep{vill14,liu13,vill11} to ultraluminous infrared 
galaxies (ULIRGs) \citep{arri14}. The properties of the ionized winds 
observed in ULIRGs seem to differ from those observed in X-ray selected 
QSOs or type 2 QSOs. For ULIRGs with a mean velocity  $\sim300$ km s$^{-1}$ 
and a mean electron density $\sim$460 cm$^{-3}$, the mass loss rate 
from ionized outflows fall in the range 
0.1$\le {\dot M_g}\le 200~M_\odot~$yr$^{-1}$
for a typical size of $\sim$1 kpc \citep{arri14}. The ionized outflows 
associated with type 2 QSOs show a larger mean velocity and the physical 
quantites varying over a larger range \citep{vill14, rupk13, liu13}. 

With a size of $\sim15$ pc (at least 100 times smaller than those found in
the extreme external galaxies referred above), the bipolar lobes in the RBZ 
of the Galactic center appear to be a mini-version of the ionized outflow, 
found in ULIRGs and type 2 QSOs.  The bipolar flows in the lobes appear to 
be sub-\Alfvenic\/~bipolar of \Alfvenic\/~Mach number ${\mathscr M}_{\rm A}\lesssim1$ 
(see Appendix B).  Then, if the SE Lobe contributes the same amount in $\dot M_{HII}$ 
as the NW Lobe, the total mass ejection rate into the lobes is 
$\dot M_{HII}\sim 0.1 M_\odot~yr^{-1}$, which falls at the lower end of 
the range for ULIRG \citep{arri14}. The total ionized mass  in the bipolar 
lobes is a few hundred times smaller than the lowest value found for outflow 
masses in the ULIRG sample \citep{arri14}. The scale of the radio lobes is 
small compared to the scale of the central molecular zone (CMZ, $\sim$400 pc 
diameter), throughout which star formation is taking place. So in the case of 
our Galaxy, we can conclude that the nuclear outflow emanates from within the 
central parsec, rather than from star formation throughout the CMZ. Thus, while 
the CMZ has been implicated in cosmic ray outflows that might have created the
Fermi Bubbles \citep{su10,croc11}, the bipolar lobes must be attributed to 
energy released within the CND.

A comparison of Fig. 3 with Figs. 4, 5 and 6 suggests that the NW Lobe 
originates from the inner cavity or the inner edge of the CND. The 
possibility that the NW ionized outflow consists of material that has 
been driven by a jet is unlikely because of the relatively large opening 
angle of the lobe, and because of the absence of a well-collimated X-ray 
jet in the NW Lobe or of a counterpart to the putative parsec-scale 
X-ray jet in the SE Lobe 
\citep{li13}. 

Another process to consider is radiation-pressure driving of the outflow.
For the individual massive stars in the central stellar cluster, it is 
indeed likely that radiation pressure drives their winds, and that 
happens very close to the photospheres of those individual stars. 
Then, the winds from all the stars, with velocities that can be 
1000 - 2000 km s$^{-1}$, coast outward and merge, creating an outflowing 
collective wind that leaves the central cluster on a scale $\sim$0.5 pc. 
The central gravity of Sgr A* and the concentrated central stars will 
substantially decelerate the collective wind. The location and orientation 
of the wings suggest that the lobes could occur as a result of 
the deflection of a wind from the central regions off the CND.

{\bf How is the outflow collimated?} The magnetic field of the CND may 
play a role, as it appears to be predominantly toroidal, or perpendicular 
to the outflow \citep{hild93}. Furthermore, if the magnetic field 
in the volume around the CND is oriented perpendicular to the Galactic plane 
or to the CND, as it appears to be in the rest of the central molecular zone  
\citep{morr06,morr14}, then the deflected, ionized wind would be guided 
to follow the vertical field lines, and would naturally assume the 
orientation of the wings. Given such a scenario, we are left with a puzzle 
of why each of the wings extends in only one direction from its interaction 
point in the CND. In a perfectly axisymmetric situation, one might expect 
the wings to be deflected in both directions out of the plane of the CND. 
If the predominant magnetic field direction in the region ovelying the 
CND is indeed perpendicular to the Galactic plane, and if the final 
direction of the outflowing wind is indeed aligned with ambient magnetic 
field then perhaps the $\sim$20\arcdeg\ tilt of the CND with respect to 
the Galactic plane strongly favors an acute deflection angle over an 
obtuse one.

{\bf Is radiation-pressure driving an important contributor to the outflow?} 
The role of nuclear radiation in driving 
ionized outflow winds has been studied in galaxy formation and AGN, 
{\it e.g.} \citep{silk98,fabi99,krum13,thom15,ishi15}. In the context 
of the Galactic center, the radiation from the central cluster needs to 
be assessed for its role in pushing out the ionized gas to a farther 
distance from the CND as well as sustaining the bipolar structure 
against the central gravity. From an hydrodynamic approach, the 
further motion of the gas is governed by the following equation 
\citep{oste74},  
\begin{eqnarray}
{dV_r \over dt} &=& - {\nabla p\over\rho} - {GM(r)\over r^2} + 
{a^2 A_{\rm d} {\mathscr L}^* \over 4 m_{\rm H} r^2c}
\end{eqnarray}
where the first term on the right is due to the pressure gradient; 
the second and the third terms are the forces per unit mass of the 
central gravity and the radiation pressure from the central cluster, 
respectively. In the following analysis, we will adopt the distribution 
of stellar mass given by \cite{genz10}, which has been described as a 
power-law outside the central cluster core ($r\ge0.25$ pc); adding 
a mass of Sgr A* $\sim4\times10^6\/ M_\odot$ at the center 
\citep{ghez05,gill09}, an enclosed mass at $r$ is 
$$M(r) \approx 1\times10^6 M_\odot\left[1.2\left(r\over {\rm pc}\right)^{1.2} + 4\right].$$  

A luminosity of ${\mathscr L}_{4.5\rm\mu m}=( 4.1\pm0.4) \times10^7 ~L_\odot$
from the Galactic center cluster has been derived from the Spitzer data 
at 4.5 $\mu$m \citep{scho14}. On other hand, \cite{mezg96} argues that 
the early B or late-type O stars produce a total ultra-violet luminosity 
in the central parsec of ${\mathscr L}_{\rm UV}\sim 7.5\times 10^7L_\odot$. 
The total luminosity from the central cluster 
${\mathscr L}^* = 7.5\pm3.5\times10^7 \/ L_\odot$ \citep{mezg96} 
is considered as an upper limit in our following analysis.

For the process of electron scattering, where $\sigma_{\rm T}$ and 
$m_{\rm p}$ are the Thomson cross section and the proton mass, 
respectively, the force due to radiation pressure 
(${\mathscr L}^*\sigma_{\rm T}\rho/4\pi m_{\rm p}  r^2 c$) 
is indeed much smaller than the central gravity. However, dust 
grains in the gas are partially charged in the environment of the 
nuclear regions \citep{fabi12} and HII regions \citep{oste74}; 
therefore, the electrical force effectively binds dust particles to 
the surrounding ionized gas. The strong coupling between the dust 
and gas will transmit the central repulsive force of radiation 
pressure to the entire ionized gas distribution. The effective 
dust cross section equivalent to $\sigma_{\rm T}$ is
$\sigma_{\rm d} = \pi\ a^2 A_{\rm d}$ for radius $a$ of a dust 
particle and an abundance $A_{\rm d}$ of dust grains by number with 
respect to hydrogen. Thus the actual radiation force is boosted by 
a factor of ${\sigma_{\rm d}/\sigma_{\rm T}}$ as compared to that 
of electron scattering. The boost factor appears to be $\sim$1000 
assumed for AGNs \citep{fabi12}, which is equivalent to $a\sim 
0.15~\mu$m and $A_{\rm d}\sim1\times10^{-12}$, a mean 
value of the ISM, {\it e.g.} \cite{ali05}. For a  dust-grain 
radius of $a=0.3~\mu$m and $A_{\rm d}=5\times10^{-12}$ used in 
modeling HII regions \citep{math67}, 
$\sigma_{\rm d} =\pi a^2 A_g \approx 1.4\times10^{-20}$ cm$^{-2}$ 
and  ${\sigma_{\rm d}/\sigma_{\rm T}}\sim2\times10^4$.
The parameters used in \cite{math67} appear to be too optimistic for 
radiation pressure, in which the dust abundance is five times higher 
than the the mean value of the ISM. In our assessment, we assume  
$a=0.3~\mu$m, a value near the peaks of size distributions for 
carbonaceous-silicate grain model of the Galaxy \citep{drai03}, 
and the mean abundance of $A_{\rm d}\approx1\times10^{-12}$ for the ISM, 
${\sigma_{\rm d}/\sigma_{\rm T}}\approx4.2\times10^3$. Thus, 
we derive the ratio of radiation force to gravity as follows, 
\begin{eqnarray}
{\mathscr X}
&\approx &1.3\times10^{-3}{\sigma_{\rm d}\over\sigma_{\rm T}}
{{\mathscr L}_{4.5\mu m}^* \over
1.2 r_{\rm pc}^{1.2}+4}
\end{eqnarray}
where ${\mathscr L}_{4.5\mu m}^*$ is the luminosity of central cluster 
in units of ${\mathscr L}_{4.5\rm\mu m}= 4.1\times10^7 ~L_\odot$ \citep{scho14}
and $r_{\rm pc}$ is the radial distance from Sgr A* in units of 1 pc. The 
Eq.(2) indicates that ${\mathscr X}\sim 1$ at $r_{\rm pc}\sim1$ for 
${\mathscr L}_{4.5\rm\mu m}^*=1$, implying that the radiation pressure 
may accelerate the ionized outflow within $r_{\rm pc}\sim1$ before 
the radiation force drops drastically compared to the gravity. Solving 
for the velocity in the hydrodynamic equation Eq.(1) assuming 
${\sigma_{\rm d}/\sigma_{\rm T}}\sim4\times10^3$, we discuss 
two models: 
A) ${\mathscr L}^*=7.5\times10^7 L_\odot$ (${\mathscr L}_{4.5\mu m}^*=1.8$) 
and the initial radial velocity of the gas $V_{\rm r}\sim0$ km s$^{-1}$ 
at $r_{\rm pc}\sim1$ in comparison to 
B) ${\mathscr L}^*=4.1\times10^7 L_\odot$ (${\mathscr L}_{4.5\mu m}^*=1$) 
and $V_{\rm r}\sim0$ km s$^{-1}$ at $r_{\rm pc}\sim0.25$.
For Model A, the ionized gas can be accelerated by the radiation forces 
to a velocity $V_{\rm r}\sim$142 km s$^{-1}$ up to $r_{\rm pc}\sim3.6$; 
beyond 3.6 pc, the gravity exceeds the radiation force and then the gas 
is decelerated. The outward velocity drops to $V_{\rm r}\sim0$ km s$^{-1}$ 
at $r_{\rm pc}\sim$ 25, giving a mean velocity $\sim$120 km s$^{-1}$. 
If the collimated outflow starts at $r_{\rm pc}\sim0.25$ as suggested 
in Model B, the primary acceleration zone by the radiation force from
the central cluster occurs in $r_{\rm pc}\sim$ 0.25 to 1; 
at $r_{\rm pc}\sim1$, the outward velocity reaches its peak
$V_{\rm r}\sim$144 km s$^{-1}$; and a maximum travel distance 
$r_{\rm pc}\sim$10 and a mean velocity $\sim$105 km s$^{-1}$ are expected. 

However, the models appear to be sensitive 
to the size of dust grains as well as their abundance in addition to the 
luminosity of the central cluster. The property of dust grains at the 
Galactic center is the key issue to nail down the role of the 
radiation pressure. The process of radiation 
pressure in driving the ionized outflow may merit further attention based 
on some known facts. With 100+ O/WR stars of $\sim$ 4 - 6 
Myr age concentrated around Sgr A* \citep{genz10,lu13}, the environment 
around Sgr A* is essentially a super HII region
mixed with the warmer dust ($T_{\rm d}\sim$ 70 -- 130 K) as well as 
the hot dust ($\sim$300 K) associated with 
the Bar and Northern Arm \citep{mezg96}.
The co-existence of dust particles along the Mini-spiral
structure \citep{geza92, aitk98, mezg96, lau13} implies that
the dust grain particles are strongly  coupled with
the ionized gas.
In addition, the OB stars outside the central cluster 
core contribute a meaningful 
luminosity besides that from the central stellar core, 
given the average 
surface density of early type stars  a few hundredth stars 
arcsec$^{-2}$ at $r>13\arcsec$ ($>$0.5 pc) \citep{buch09, stos15}. 
Furthermore, the directions of the SE and NW Wings' extension appear 
not to point back to Sgr A*, implying that 
the radiation from the central cluster  may  play an appreciable role 
at $r\sim0.25$ to 1 pc in driving out the ionized gas from 
the CND to the radio lobes.

Finally, major events from Sgr A* happen episodically along with
large flares that likely add a substantial flux of high-energy 
photons  up to $\sim10^{41-42}$ erg s$^{-1}$ at the Galactic center. 
The 100-year events suggested by the front of 
fluorescent X-rays  propagating away from Sgr A*, as shown {\it e.g.} in 
\cite{pont10} and \cite{clav13}, provide evidence for such past activity.  
Extraodinary X-ray flares are also expected from the statistical analysis  
of the flux-density fluctuations observed in the near infrared \citep{witz12}.

\section{Summary and Conclusion}

Using the Jansky VLA in the B and C arrays, we have observed Sgr A 
using the broadband (2 GHz) continuum mode at 5.5 GHz covering the
central 13\arcmin~(30 pc) region of the RBZ at the Galactic center.
Using the CASA multi-scale and multi-frequency-synthesis clean 
algorithm, we have constructed a sensitive image, achieving an rms 
noise level of 8 $\mu$Jy beam$^{-1}$ and a dynamic range of 100,000:1.
The high dynamic range allows us to  unambiguously  distinguish critical 
emission features in the radio. The new broadband VLA image shows a 
vast number of filamentary structures in the Sgr A complex surrounding 
the supermassive black hole, Sgr A*. The image 
provides unprecedented detail on several prominent emission features, 
including the NW and SE ``Wings'' of the Mini-spiral in Sgr A West, a 
$\Sigma$-shaped band of emission -- the Sigma Front -- that lies within 
the shell of the Sgr A East supernova remnant, and the radio blobs and 
streamers in the radio lobes situated at positive and negative Galactic 
latitudes from Sgr A*.

We have compared the structures of the radio sources at 6 cm with molecular
line images and X-ray images, finding that the Sigma Front appears to 
mimic the shape and orientation of the high-density molecular gas in 
the adjacent CND. A simple geometrical model is examined to demonstrate 
that the Sigma Front can be created by the impact and reflection 
of the blast waves from Sgr A East.  This model constrains the location
of the SN explosion site to be at its projected distance of 2 parsec east of 
Sgr A* and less than 3.3 pc behind the sky plane containing Sgr A*. 
The observation that Sgr A West is seen in absorption against Sgr A East 
at 90 cm is consistent with this model.

The new 5.5-GHz image provides considerable detail on the radio counterpart 
of the bipolar structure observed in X-rays,  centered on Sgr A* 
\citep{bag03,morr03,pont15}.  With a detailed comparison of the radio 
image with the HST/NICMOS P$\alpha$ image, we find that most of the 
radio continuum emission in the NW bipolar lobe is thermal free-free emission.

\acknowledgments
{We thank  S. Mart\'in for providing the SMA CN image.
We are grateful to R. V. Urvashi for assistance with and advice on
the use of CASA. The Very Large
Array (VLA) is operated by the National Radio Astronomy Observatory
(NRAO). The NRAO is a facility of the National Science Foundation
operated under cooperative agreement by Associated Universities, Inc.
The research has made use of NASA's Astrophysics Data System.
}
\appendix
\section{A. Data Reductions}

\subsection{Basic Calibrations}

The data reduction was carried out using the software package of 
CASA-Common Astronomy Software Applications ({\bf http://casa.nrao.edu})
of the NRAO. The standard 
calibration procedure for continuum VLA data was applied. J1733-1304 
(NRAO 530) was used for complex gain calibrations. The flux density 
scale was calibrated using the primary calibrators, either 3C 286 
(J1331+3030) or 3 C48 (J0137+3309). Corrections for the bandpass shape 
of each baseband and the delay across the 2GHz band were determined 
based on the primary calibrator data.

\subsection{Removal of Time Variation of Sgr A*}

We investigated the time variation of the flux density from Sgr A* by 
creating lightcurves with the vector averaged visibilities for all 
baselines in the range between 60 and 300 k$\lambda$ in 5-min time bins, 
phased at the position of Sgr A* using AIPS task {\it DFTPL}. The high-S/N 
central 85\% of the data channels in each of the 16 spectral subbands are 
used in the calculations of the lightcurves. The variabilities of Sgr A*, 
defined numerically as $2\left({S_{\rm Max} - S_{\rm Min}}\right)
/\left({S_{\rm Max} + S_{\rm Min}}\right)$, where $S_{\rm Max}$ and 
$S_{\rm Min}$ are the maximum and minimum flux densities, respectively, 
measured during the observation, are up to 20\% within each of the four 
observing dates, and 50\% between the dates. The effect of flux density 
variations of Sgr A* on the imaging quality is discussed by \cite{zhao91}. 
We have developed a CASA/AIPS procedure to remove the time variation 
of Sgr A* prior to further data reduction in order to achieve a high-dynamic 
range image. Given an epoch of observation, the corresponding data set 
is broken down into numerous subsets of UV data, each consisting of data 
within a 5-min time interval. The variable component at the position 
Sgr A* in each subset of data was modeled using a CASA module 
{\it Addcomponent} assuming a Gaussian shape of Sgr A*. 
The major and minor axis sizes as well as the position angle of the major 
axis were determined from the scattering shape of Sgr A*, 
$\theta_{\rm Maj}=0.042\arcsec$, $\theta_{\rm Min}=0.023\arcsec$, 
PA=80\arcdeg, {\it e.g.}, \cite{lo98} and \cite{bow04}, and the flux 
density for a given 5-min time interval was determined from the lightcurve 
using {\it DFTPL}. Then, the variable component models were Fourier-transformed 
({\it FT}) to the UV domain and attached to their corresponding UV subsets. 
The variable component Sgr A* are subtracted from the UV subsets with 
{\it UVSUB} to create the residuals without Sgr A*. With CASA task 
{\it CONCAT}, we concatenate the subsets back to the four observing-epoch 
sets. Finally, the Gaussian source at the position of Sgr A* with a constant value of 
0.8 Jy (the average flux density) is added back to the residual data sets 
with {\it FT} and {\it UVSUB} in CASA. Therefore, the time-variable 
component at the position of Sgr A* has been removed from the visibility 
data. Fig. A1 outlines the procedure of correcting for the time-variable 
flux density of Sgr A* in CASA. The relevant CASA tasks and modules are 
scripted in a Python program.

\begin{figure}[t]
\hskip 60mm
\includegraphics[angle=0,width=120mm]{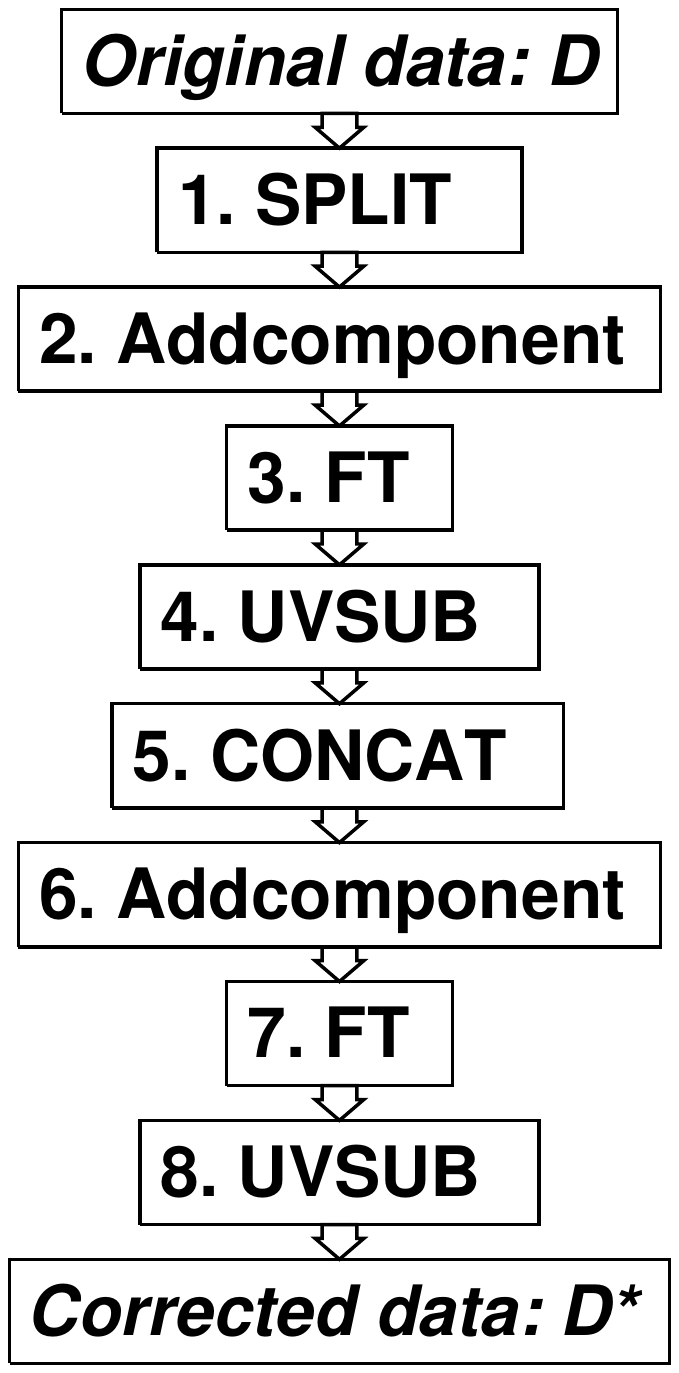}
\vskip -65mm
\caption{A flow chart of the procedure to remove the time variable component of Sgr A*,
in eight steps. The names of the CASA tasks and modules used in each step
are given in each of the corresponding boxes. The details in each step are described
as follows: \\
{\bf 1.~SPLIT} - {\it split an original data set D into subsets d(t);}\\
{\bf 2.~Addcomponent} - {\it create a model of Sgr A* at time t;}\\
{\bf 3.~FT} - {\it FFT the Sgr A* model into the UV plane;}\\
{\bf 4.~UVSUB} - {\it subtract Sgr A* from  d(t) to obtain a residual data rd(t);}\\
{\bf 5.~CONCAT} - {\it concatenate the rd(t) files together to assemble RD;}\\
{\bf 6.~Addcomponent} - {\it create a Sgr A* model with its mean flux;}\\
{\bf 7.~FT} - {\it FFT the mean flux model into the UV plane, attaching it to RD;}\\
{\bf 8.~UVSUB} - {\it add Sgr A* back to RD to produce D*.}\\
The time variation of the flux-density from Sgr A* has been
removed in the final dataset {\it D*} produced from this procedure.\\
~ \\
}
\end{figure}

\begin{table}[b]
\vskip 7mm
\setlength{\tabcolsep}{0.7mm}
\caption{A list of the prominent sources in the radio bright zone (RBZ)}
\begin{tabular}{llll}
\hline\hline \\
{Source}&
{Name}&
{Description} &
{Ref.}\\
\hline
J174542.00$-$290018.0~~&Sgr A Complex&Complex of numerous radio sources in the RBZ in an
                                     angular diameter of 13\arcmin\/&                                     1,2,3\\
J174540.04$-$290028.1&Sgr A*&Compact radio source, associated with the supermassive black hole (SMBH)&  4,5,6\\
J174542.88$-$290018.2&Sgr A East&Shell of radio emission 3.3\arcmin$\times$2.3\arcmin (35\arcdeg),
                                possible supernova remnant (SNR)&                                          7,8\\
J174543.65$-$290040.0&Sigma Front& $\Sigma$-shaped radio feature, consisting of four segments, located within Sgr A East    &9\\
J174540.04$-$290029.7&Sgr A West&Radio structure, 40\arcsec$\times$75\arcsec, characterized as the Mini-spiral
                                  &7,8,13\\
J174540.04$-$290031.4&Bar&Central bright part of  Mini-spiral, 8\arcsec$\times$4\arcsec, $\sim90$\arcdeg\ &8\\
J174540.43$-$290013.2&Northern Arm&Feature located NE of Sgr A*, extended $\sim40$\arcsec\ to north         &8\\
J174541.04$-$290033.4&Eastern Arm&Feature located SE of Sgr A*, extended $\sim40$\arcsec\ to northeast      &8\\
J174538.74$-$290043.4&Western Arc&Feature located SW of Sgr A*, arc extended $\sim60$\arcsec           &8\\
J174540.05$-$290027.3&CND        &Circum-nuclear disk, characherized by dense molecular ring 
                                              80\arcsec$\times$40\arcsec, $\sim25$\arcdeg\            &10,11,12\\
J174537.82$-$285902.1&NW Wing&Northwestern wing of the Mini-spiral, extended $\sim2\arcmin$              &9\\
J174537.99$-$285911.0&NW Streak&Short filament in the NW Wing, counterpart of
                                      CXOUGCJ174538.0$-$285911                                       &1,3,14,15\\
J174526.00$-$285915.0&NW Lobe&Complex extended emission region (6.3\arcmin$\times$3.2\arcmin, $-70$\arcdeg),
                            located NW of Sgr A*                                                           &1,9\\
J174521.65$-$285826.2&Smoke Rings&Three ring-like emission features, located in NW Lobe                  &9\\
J174545.53$-$285828.7&Cannonball&Radio counterpart of the X-ray Cannonball, 
                                                                            pulsar wind nebula (PWN)&14,15,16\\
J174541.74$-$285856.1&Northern Filaments&Radio counterpart of the NuSTAR source 
                                                                               in hard X-ray, G359.97$-$0.038&17\\
J174541.73$-$285816.6&Northern Curls&Source I1/I2, a bundle of filaments, located at northern 
                                                  edge of Sgr A East                                    &1,3,18\\
J174536.84$-$290039.3&M &Head-tail source, head corresponding to CXOUGCJ174536.9$-$290039
                                                                                                        &1,9,14\\
J174528.60$-$290421.9&SW Shell&Shell-like source, 6\arcsec$\times$5\arcsec\ (28\arcdeg)                   &9\\
J174541.71$-$290126.0&SE Wing &Southeastern wing of Mini-spiral, extended $\sim2\arcmin$             &9\\
J174545.00$-$290300.0&SE Lobe &Radio counterpart of the SE X-ray lobe                                    &9\\
J174547.82$-$290305.6&SE Blobs&Amorphous radio blobs, located $\sim3\arcmin$ SE of Sgr A*                    &9\\
J174547.22$-$290253.6&SE Cone&Cone shaped emission feature,  apex pointing away from Sgr A*            &9\\            
J174553.21$-$290227.9&SE Trunk&Trunk emission, 115\arcsec$\times$15\arcsec\ ($\sim152\arcdeg$), 
                                                                                located SE of  Sgr A East&1,2,9\\
J174540.97$-$290437.3&Southern Curls&Sources E and F,  bundle of filaments, 
                    located $\sim4$\arcmin\ south of Sgr A*                                              &18,19\\
J174600.73$-$290048.3&Streaks&Group of filaments, $\sim2\arcmin$ in length, located east of Sgr A East   &1,3\\
J174540.00$-$290000.0&CMZ&Central molecular zone, strong gas concentration in central 
                                                                                     $\sim3$\arcdeg (400 pc)&20\\
\hline
\end{tabular}\\
\begin{tabular}{p{0.95\textwidth}}
{\footnotesize \it 1. Yusef-Zadeh \& Morris (1987);          
2. Pedlar et al.\ (1989);                 
3. Yusef-Zadeh, Hewitt \& Cotton (2004);  
4. Balick \& Brown (1974);                
5. Reid (2009);                           
6. Genzel (2010);                         
7. Ekers et al. (1975);                   
8. Ekers et al. (1983);                   
9. Zhao, Morris \& Goss (this paper);     
10. G\"usten et al. (1987)                
11. Jackson et al. (1993)                 
12. Mart\'in et al.\ (2012)               
13. Lo \& Claussen (1983);                
14. Muno et al.\ (2009);                  
15. Zhao, Morris \& Goss (2013);          
16. Park et al.\ (2005);                  
17. Nynka et al.\ (2015);                 
18. Morris, Zhao \& Goss (2014);          
19. Ho et al.\ (1985);                    
20. Morris \& Serabyn (1996).             
}\\
\end{tabular}
\end{table}

\subsection{Self-Calibrations}

The residual errors were further corrected in two steps
with self-calibration techniques using a clean component
model created from the MS-MFS clean process.
First, we calculated gain corrections using the CASA task 
{\it GAINCAL} with phase only for a few cycles until
the phase errors in the visibility data were minimized.
Then, we corrected for amplitude using {\it GAINCAL} with the 
{\it calmode}='{\bf ap}', employing the gain table produced 
from the last application of the phase-only step.
Thus, the gain corrections calculated from the second step were
only for amplitudes. Again, a few cycles for amplitude
corrections were carried out until the corrections converged.

\subsection{RFI Rejection}

Rejection of radio frequency interference (RFI)
was the most time-consuming task, but was critical 
to achieve a high-dynamic-range image.
The procedure of RFI rejection occurred in three
steps. First, the interactive tool {\it PLOTMS} was 
used to identify and then flag obvious RFI from the data.
RFI in the calibrators with simple structure can be easily 
identified and rejected. More subtle RFI
was then edited using the CASA modules {\it TFCROP} and
{\it RFFLAG} for the calibrated data of the target source with
complex visibility structure. Finally,
during each of the self-calibration cycles, the low-level
RFI was identified and flagged using {\it PLOTMS} for
the residual visibilities (the calibrated
visibility subtracted with the clean component model)
for each baseline.

\subsection{MS-MFS Imaging}
After eliminating the RFIs and calibrations,
we constructed an image with the four-epoch  broadband data sets.
The dirty image was cleaned using the MS-MFS algorithm
\citep{rau11} by fitting the amplitude change across the 2 GHz
band with  the first two terms in a  Taylor
expansion of the intensity spectrum at each pixel. The first term, 
the zeroth order Taylor expansion ({\it tt0}),
corresponds to the Stokes-I image at the reference frequency
(5.5 GHz), the center frequency in the 2-GHz band.
The spectral index image, $\alpha-$image, is generated from the
second term in the first-order Taylor expansion ({\it tt1}).
Two images were then  produced. One is constructed
with
all the C and B (C+B) array data with a  weighting of $robust=0$,
to image overall emission structure.
The rms noise of the Stokes-I image prior to the primary-beam corrections
is 8 $\mu$Jy beam$^{-1}$, determined from areas with no sources. 
The second image was made from the
B array data only
with uniform weighting ($robust=-2$) in order to detect compact sources.
The rms noise of this image is 15 $\mu$Jy beam$^{-1}$.

\subsection{Primary-Beam Correction}
We have corrected for the primary beam attenuation up to the level of 10\% of the 
value at the pointing center, corresponing to a diameter of 13\arcmin.1 for
the field of view in the final image.

\subsection{List of Prominent Radio Sources}

In Table A1, we list the prominent radio sources that are discussed in the text.
The corresponding J2000 coordinates of the sources and their nomenclatures adopted 
or created in this 
paper are given in Columns 1 and 2, respectively.
A discription for each of the sources are supplied in Column 3 along with 
relevant references (Column 4).   

\section{B. Power of The Bipolar Outflow}
The well-defined NW radio lobe, consisting primarily of free-free 
emission, might represent a significant feedback from the 
region surrounding Sgr A* via an ionized  wind driven by 
the overwhelming radiation pressure from the central cluster; and the wind
is possibly collimated  by the local poloidal magnetic 
field anchored on the CND. In the following, we will assess the power of 
the possible bipolar ionized outflow, utilizing the NW radio 
lobe as an exercise. The power transported from a non-relativistic 
outflow wind can be estimated,
\begin{equation}
{\mathscr P}_{w} = \pi r_{\rm w}^2 v_{\rm w} {\mathscr U}_{\rm E},
\end{equation}
\noindent where $r_w$ is the radius of a cross section of the collimated
outflow wind with its velocity $v_{\rm w}$ and volume
${\mathscr V}_{\rm w}$; and
the energy density including  kinetic ($E_{\rm K}$),
internal ($E_{\rm in}$) and magnetic field ($E_{\rm B}$) is:
\begin{equation} 
{\mathscr U}_{\rm E} = {\displaystyle E_{\rm K} + E_{\rm in} + E_{\rm B} \over
         \displaystyle {\mathscr V}_{\rm w} }.
\end{equation}

The free-free ($ff$) flux density from the ionized outflow can be expressed as  
$S_{ff}=\zeta_{ff}S_{5.5 \rm GHz}$ where $\zeta_{ff}\lesssim1$ is a dilution 
factor owing to contamination from the diffuse non-thermal synchrotron 
component \citep{pedl89} and the non-outflow f-f emission from foreground/background
nebulae; 
and $S_{5.5 \rm GHz}$ is the total flux density from a lobe. 
With the radio measurements, we are able to assess the physical quantities 
of the ionized outflow.  
The mean electron density ($n_{\rm e}$) and HII mass ($M_{\rm HII}= 
m_{\rm p}n_{\rm e}{\mathscr V}f_{\rm V}$) 
can be estimated with the formula deduced from \cite{mezg67}
using cylindrical geometry 
(${\mathscr V}={\displaystyle \pi\over \displaystyle 4}\theta^3D^3$) 
with volume 
filling factor of the ionized gas $f_{\rm V}$ and proton mass $m_{\rm p}$:
\begin{eqnarray}
n_{\rm e} &\approx& 5.4\times10^2~{\rm cm}^{-3}f_{\rm V}^{-1/2}\zeta^{1/2}_{ff}\left(T_{\rm e}\over6000 {\rm K}\right)^{0.175}
\left(S_{5.5\rm GHz}\over {\rm Jy}\right)^{0.5}\left(D\over {\rm kpc}\right)^{-0.5}
\left(\theta_G \over {\rm arc~min} \right)^{-1.5}, \\
M_{\rm HII} &\approx& 0.45~M_\odot~ f_{\rm V}^{1/2}\zeta^{1/2}_{ff}\left(T_{\rm e}\over
6000 {\rm K}\right)^{0.175}
\left(S_{\rm 5.5GHz}\over Jy\right)^{0.5}\left(D\over {\rm kpc}\right)^{2.5}
\left(\theta_{\rm G} \over {\rm arc~min} \right)^{1.5},
\end{eqnarray}

\noindent Thus, we have 
$n_{\rm e} \sim 109 f_{\rm V}^{-1/2} \zeta^{1/2}_{ff} ~{\rm cm}^{-3}$
and $M_{\rm HII}\sim 2.4\times10^3 f_{\rm V}^{1/2} \zeta^{1/2}_{ff} ~M_\odot$
with the measured quantities from the VLA observations
at 5.5 GHz:  $S_{5.5\rm GHz}\sim17$ 
Jy, $\theta_{\rm G}={\displaystyle1\over\displaystyle1.201}\sqrt{\theta_{\rm maj}\times\theta_{\rm min}}\sim{\displaystyle1\over\displaystyle1.201}\sqrt{6.3\arcmin\times3.2\arcmin}$, $T_{\rm e}\sim6\times10^3$ K
and $D=8$ kpc. The expansion 
time scale of the NW Lobe can be estimated  
assuming a constant velocity 
$v_{\rm w}$ along the rotational axis of the CND, which  has an inclination angle
of $i\sim60$\arcdeg \citep{gust87,zhao10}, 
\begin{eqnarray}
t_{\rm w} \sim {\displaystyle 
\theta_{\rm maj}D\over \displaystyle v_{\rm w} {\rm sin}(i)} 
\approx8.7\times10^4~{\rm yr}~{\mathscr M}_{\rm A}^{-1}\left(\theta_{\rm maj}\over 6.3\arcmin\right)
\left(D\over 8~{\rm kpc}\right)
\left(B\over 1~{\rm mG}\right)^{-1} \left(\displaystyle n_{\rm i}\over \displaystyle 100~{\rm cm^{-3}}\right)^{0.5},
\end{eqnarray}
given the outflow velocity being scaled with \Alfvenic Mach number
${\mathscr M}_{\rm A}=v_{\rm w}/ v_{\rm A}$ with an \Alfvenic velocity:

\begin{equation} 
v_{\rm A} = {B\over \sqrt{4\pi\rho}} \approx 190~{\rm km~s^{-1}} 
\left(\displaystyle B\over \displaystyle 1~{\rm mG}\right)
\left(\displaystyle n_{\rm i}\over \displaystyle 100~{\rm cm^{-3}}\right)^{-0.5}.
\end{equation}
Taking the ion mass $M_{\rm ion} = \mu M_{\rm HII}$ and $\mu=1.3$ assuming
solar abundances,
the mass loss rate via the NW Lobe is
\begin{eqnarray}
{\dot M} &\approx& 3.6\times10^{-2}~M_\odot~{\rm yr}^{-1}
{\mathscr M}_{\rm A}
f_{\rm V}^{1/2} \zeta^{1/2}_{ff} \nonumber \\
  & & 
\left(T_e \over 6000 {\rm K}\right)^{0.175}
\left(S_{5.5{\rm GHz}}\over 17 \/ {\rm Jy}\right)^{0.5}
\left(D\over 8~{\rm kpc}\right)^{1.5} 
\left(\theta_{\rm maj}\over 6.3\arcmin\right)^{0.5}
\left(\theta_{\rm min}\over 3.2\arcmin\right)^{1.5}
\left(B\over 1~{\rm mG}\right) \left(\displaystyle n_{\rm i}\over \displaystyle 100~{\rm cm^{-3}}\right)^{-0.5}.
\end{eqnarray}
The kinetic energy of bulk motion ($E_{\rm K}$) and internal enery ($E_{\rm in}$) 
can be derived as follows:

\begin{eqnarray}
E_{\rm K}& = &{1\over2} \mu\/M_{\rm HII} v_{\rm w}^2 \nonumber\\
     & \approx &1.2\times10^{51}~{\rm ergs}~
{\mathscr M}_{\rm A}^2 f_{\rm V}^{1/2} \zeta^{1/2}_{ff} \nonumber\\
     & & \left(T_{\rm e} \over 6000~{\rm K}\right)^{0.175}
\left(S_{5.5{\rm GHz}}\over 17 \/ {\rm Jy}\right)^{0.5}
\left(D\over 8~{\rm kpc}\right)^{2.5} 
\left(\theta_{\rm maj}\times \theta_{\rm min}
\over  6.3\arcmin \times 3.2\arcmin\right)^{1.5}
\left(B\over 1~{\rm mG}\right)^2 \left(\displaystyle n_{\rm i}\over 
\displaystyle 100~{\rm cm^{-3}}\right)^{-1},  \\
  & & \nonumber \\
E_{\rm in} &=& {\mathscr V}f_{\rm V}{P_{\rm th}\over \gamma-1}\nonumber\\
           &\approx& 7.0\times10^{48}~{\rm ergs}~f_{\rm V}\left(\theta_{\rm maj}
\over 6.3\arcmin\right)  
\left(\theta_{\rm min}\over 3.2\arcmin\right)^2
\left(D\over8~{\rm kpc}\right)^3
\left(T_{\rm e}\over 6000~{\rm K}\right)
\left(n_{\rm i}\over100~\rm cm^{-3}\right),  
\end{eqnarray}
where we take the thermal pressure $P_{\rm th}=2n_ekT_e$
and ${\mathscr V}\approx{\displaystyle \pi\over \displaystyle 4}\theta_{\rm maj}\theta_{\rm min}^2D^3$, 
adiabatic coefficient $\gamma=5/3$ and 
assuming a cylindrical geometry for
the outflow. For a typical magnetic field strength $B\sim$1 mG
with a volume filling factor $f_{\rm B}$ of the magnetized region,
the magnetic energy stored in the flow is

\begin{eqnarray}
E_{\rm B} &=&{\mathscr V} f_{\rm B}{B^2\over 8\pi} \nonumber \\
          &\approx&
7.4\times10^{50}~{\rm ergs}~f_{\rm B}\left(\theta_{\rm maj}\over 6.3\arcmin\right)  
\left(\theta_{\rm min}\over 3.2\arcmin\right)^2 
\left(D\over8~{\rm kpc}\right)^3
\left(B\over1~{\rm mG}\right)^2. 
\end{eqnarray}

\noindent Then, from the physical quantities of the NW lobe determined
from the VLA observations at 5.5 GHz and other observations,
we estimate the kinetic energy 
$E_{\rm  K} \sim 1.2\times10^{51}~{\rm ergs}~f_{\rm V}^{1/2} 
\zeta^{1/2}_{ff} {\mathscr M}_{\rm A}^2$.
The Model B  discussed in section 4.2 gives the mean outflow velocity
105 km s$^{-1}$, implying a mean \Alfvenic Mach number ${\mathscr M}_{\rm A}\sim0.6$
for the NW lobe.
For $f_{\rm V}\sim1$, $f_{\rm B}\sim1$
and $\zeta\sim1$, we estimate $E_{\rm K} \sim 4\times10^{50}$~ergs,
$E_{\rm in} \sim 4\times10^{48}$~ergs, and $E_{\rm B} \sim 7\times10^{50}$~ergs.
The internal energy appears to be two orders in magnitude below 
the others.  A total energy budget appears to be $E_{\rm tot}\sim1\times10^{51}$ ergs.
Thus, the total power of the NW outflow wind with a travel distance $l$ approximately 
equals to 
\begin{eqnarray}
{\mathscr P}_{\rm w} &\approx& 2\times10^{38} {\rm erg~s^{-1}}~
\left(v_{\rm w}\over 100 {\rm km s^{-1}}\right)
\left(l\over 15 {\rm pc}\right)^{-1}\left(E_{\rm tot}\over 10^{51} {\rm ergs }\right)
\end{eqnarray}
The total power transported into the NW and SE 
lobes is about three orders of  magnitude  less than the 
stellar luminosity emerging from the central few parsecs.
If the ionized bipolar outflow is powered by the  
radiation produced by the central star cluster 
(${\mathscr L^*}\approx 4.1\times
10^7 L_\odot$, and also see section 4.2), 
the efficiency of powering the outflow
by the radiation luminosity 
is $\varepsilon\approx {\displaystyle 2{\mathscr P_{\rm w}}\over \displaystyle
{\mathscr L^*}} \lesssim 0.3 \% $.

\end{document}